%Paper: hep-th/9410188
%From: chered@math.unc.edu (Ivan Cherednik)
%Date: Tue, 25 Oct 1994 13:20:31 +0500

%&amstex
%DIFFERENCE-ELLIPTIC OPERATORS AND ROOT SYSTEMS, Ivan Cherednik

\def\spaces{\space\space\space\space\space\space\space\space\space\space}
\def\spacess{\message{\spaces\spaces\spaces\spaces\spaces\spaces\spaces}}
\spacess
\spacess
\message{Annals of Mathematics Style: Current Version: 1.1. June 10, 1992}
\spacess
\spacess
%%%%%%%%%%%%%%%%%%%%%%%%%%%%%%%%%%%%%%%%%%%%%%%%%%%%%%%%

\catcode`\@=11

\hyphenation{acad-e-my acad-e-mies af-ter-thought anom-aly anom-alies
an-ti-deriv-a-tive an-tin-o-my an-tin-o-mies apoth-e-o-ses apoth-e-o-sis
ap-pen-dix ar-che-typ-al as-sign-a-ble as-sist-ant-ship as-ymp-tot-ic
asyn-chro-nous at-trib-uted at-trib-ut-able bank-rupt bank-rupt-cy
bi-dif-fer-en-tial blue-print busier busiest cat-a-stroph-ic
cat-a-stroph-i-cally con-gress cross-hatched data-base de-fin-i-tive
de-riv-a-tive dis-trib-ute dri-ver dri-vers eco-nom-ics econ-o-mist
elit-ist equi-vari-ant ex-quis-ite ex-tra-or-di-nary flow-chart
for-mi-da-ble forth-right friv-o-lous ge-o-des-ic ge-o-det-ic geo-met-ric
griev-ance griev-ous griev-ous-ly hexa-dec-i-mal ho-lo-no-my ho-mo-thetic
ideals idio-syn-crasy in-fin-ite-ly in-fin-i-tes-i-mal ir-rev-o-ca-ble
key-stroke lam-en-ta-ble light-weight mal-a-prop-ism man-u-script
mar-gin-al meta-bol-ic me-tab-o-lism meta-lan-guage me-trop-o-lis
met-ro-pol-i-tan mi-nut-est mol-e-cule mono-chrome mono-pole mo-nop-oly
mono-spline mo-not-o-nous mul-ti-fac-eted mul-ti-plic-able non-euclid-ean
non-iso-mor-phic non-smooth par-a-digm par-a-bol-ic pa-rab-o-loid
pa-ram-e-trize para-mount pen-ta-gon phe-nom-e-non post-script pre-am-ble
pro-ce-dur-al pro-hib-i-tive pro-hib-i-tive-ly pseu-do-dif-fer-en-tial
pseu-do-fi-nite pseu-do-nym qua-drat-ics quad-ra-ture qua-si-smooth
qua-si-sta-tion-ary qua-si-tri-an-gu-lar quin-tes-sence quin-tes-sen-tial
re-arrange-ment rec-tan-gle ret-ri-bu-tion retro-fit retro-fit-ted
right-eous right-eous-ness ro-bot ro-bot-ics sched-ul-ing se-mes-ter
semi-def-i-nite semi-ho-mo-thet-ic set-up se-vere-ly side-step sov-er-eign
spe-cious spher-oid spher-oid-al star-tling star-tling-ly
sta-tis-tics sto-chas-tic straight-est strange-ness strat-a-gem strong-hold
sum-ma-ble symp-to-matic syn-chro-nous topo-graph-i-cal tra-vers-a-ble
tra-ver-sal tra-ver-sals treach-ery turn-around un-at-tached un-err-ing-ly
white-space wide-spread wing-spread wretch-ed wretch-ed-ly Brown-ian
Eng-lish Euler-ian Feb-ru-ary Gauss-ian Grothen-dieck Hamil-ton-ian
Her-mit-ian Jan-u-ary Japan-ese Kor-te-weg Le-gendre Lip-schitz
Lip-schitz-ian Mar-kov-ian Noe-ther-ian No-vem-ber Rie-mann-ian
Schwarz-schild Sep-tem-ber Za-mo-lod-chi-kov Knizh-nik quan-tum Op-dam
Mac-do-nald Ca-lo-ge-ro Su-ther-land Mo-ser Ol-sha-net-sky  Pe-re-lo-mov }

\Invalid@\nofrills
\Invalid@\usualspace
\newif\ifnofrills@
\def\nofrills@#1#2{\relaxnext@
  \DN@{\ifx\next\nofrills
    \nofrills@true\let#2\relax\DN@\nofrills{\nextii@}%
  \else
    \nofrills@false\def#2{#1}\let\next@\nextii@\fi
\next@}}
\def\usualspace@#1{\ifnofrills@\def\usualspace{#1}\fi}
\def\addto#1#2{\csname \expandafter\eat@\string#1@\endcsname
  \expandafter{\the\csname \expandafter\eat@\string#1@\endcsname#2}}
\newdimen\bigsize@
\def\big@#1#2{{\hbox{$\left#2\vcenter to#1\bigsize@{}%
  \right.\nulldelimiterspace\z@\m@th$}}}
\def\big{\big@\@ne}
\def\Big{\big@{1.5}}
\def\bigg{\big@\tw@}
\def\Bigg{\big@{2.5}}
\def\raggedcenter@{\leftskip\z@ plus.4\hsize \rightskip\leftskip
 \parfillskip\z@ \parindent\z@ \spaceskip.3333em \xspaceskip.5em
 \pretolerance9999\tolerance9999 \exhyphenpenalty\@M
 \hyphenpenalty\@M \let\\\linebreak}
\def\upperspecialchars{\def\ss{SS}\let\i=I\let\j=J\let\ae\AE\let\oe\OE
  \let\o\O\let\aa\AA\let\l\L}
\def\uppercasetext@#1{%
  {\spaceskip1.2\fontdimen2\the\font plus1.2\fontdimen3\the\font
   \upperspecialchars\uctext@#1$\m@th\aftergroup\eat@$}}
\def\uctext@#1$#2${\endash@#1-\endash@$#2$\uctext@}
\def\endash@#1-#2\endash@{%
\uppercase{#1}\if\notempty{#2}--\endash@#2\endash@\fi}
\def\runaway@#1{\DN@{#1}\ifx\envir@\next@
  \Err@{You seem to have a missing or misspelled \string\end#1 ...}%
  \let\envir@\empty\fi}
\newif\iftemp@
\def\notempty#1{TT\fi\def\test@{#1}\ifx\test@\empty\temp@false
  \else\temp@true\fi \iftemp@}

%\comment%%% remove
\font@\tensmc=cmcsc10
\font@\sevenex=cmex7
\font@\sevenit=cmti7
\font@\eightrm=cmr8 % preloaded in plain.tex
\font@\sixrm=cmr6 % preloaded in plain.tex
\font@\eighti=cmmi8     \skewchar\eighti='177 % preloaded
\font@\sixi=cmmi6       \skewchar\sixi='177   % preloaded
\font@\eightsy=cmsy8    \skewchar\eightsy='60 % preloaded
\font@\sixsy=cmsy6      \skewchar\sixsy='60   % preloaded
\font@\eightex=cmex8 %
\font@\eightbf=cmbx8 % preloaded in plain.tex
\font@\sixbf=cmbx6   % preloaded in plain.tex
\font@\eightit=cmti8 % preloaded in plain.tex
\font@\eightsl=cmsl8 % preloaded in plain.tex
\font@\eightsmc=cmcsc10
\font@\eighttt=cmtt8 % preloaded in plain.tex
%\font@\ninerm=cmr9
%\font@\ninei=cmmi9    \skewchar\ninei='177
%\font@\ninesy=cmsy9   \skewchar\ninesy='60
%\font@\nineex=cmex9
%\font@\ninebf=cmbx9
%\font@\nineit=cmti9
%\font@\ninesl=cmsl9
%\font@\ninesmc=cmcsc9
%\font@\ninemsa=msam9
%\font@\ninemsb=msbm9
%\font@\nineeufm=eufm9
%\endcomment%%%

\loadmsam
\loadmsbm
\loadeufm
\UseAMSsymbols

\def\penaltyandskip@#1#2{\relax\ifdim\lastskip<#2\relax\removelastskip
      \ifnum#1=\z@\else\penalty@#1\relax\fi\vskip#2%
  \else\ifnum#1=\z@\else\penalty@#1\relax\fi\fi}
\def\nobreak{\penalty\@M
  \ifvmode\def\penalty@{\let\penalty@\penalty\count@@@}%
  \everypar{\let\penalty@\penalty\everypar{}}\fi}
\let\penalty@\penalty

\def\block{\RIfMIfI@\nondmatherr@\block\fi
       \else\ifvmode\vskip\abovedisplayskip\noindent\fi
        $$\def\endblock{\par\egroup$$}\fi
  \vbox\bgroup\advance\hsize-2\indenti\noindent}
\def\endblock{\par\egroup}

\def\logo@{\baselineskip2pc \hbox to\hsize{\hfil\eightpoint Typeset by
 \AmSTeX}}

%%%%%%%%%%%%%%%%%%%%%%%%%%%%%%%%%%%%%%%%%%%%%%%%%%%%%%%%%%%%%%%
%% Macros for Annals of Mathematics written by Amy Hendrickson
%% TeXnology Inc, Brookline, MA
%% 617 738-8029, amyh@ai.mit.edu
%%%%%%%%%%%%%%%%%%%%%%%%%%%%%%%%%%%%%%%%%%%%%%%%%%%%%%%%%%%%%%%

%% This file includes:
%% 1) Font declarations,
%% 2) Page set up,
%% 3) Title page
%% 4) Section heads,
%% 5) Equation macros, autonumbering equations, etc.,
%% 6) Figure and Table Captions,
%% 7) End matter macros: Bibliography, Appendix, etc.,
%% 8) Footnotes,
%% 9) Theorem type environments
%% 10) Cross-referencing
%% 11) Listing
%% 12) Article and Journal Table of Contents

%%%%%%%%%%%%%%%%%%%%%%%%%%%%%%%%%%%
%% 1) Font declarations,
% Computer Modern fonts

% Small Caps
\font\elevensc=cmcsc10 scaled\magstephalf
\font\tensc=cmcsc10

\font\eightsc=cmcsc10 scaled800

\font\elevenrm=cmr10 scaled \magstephalf%!!!
\font\ninerm=cmr9
\font\eightrm=cmr8
\font\sixrm=cmr6
\font\fiverm=cmr5

\font\eleveni=cmmi10 scaled\magstephalf
\font\ninei=cmmi9
\font\eighti=cmmi8
\font\sixi=cmmi6
\font\fivei=cmmi5
\skewchar\ninei='177 \skewchar\eighti='177 \skewchar\sixi='177
\skewchar\eleveni='177

\font\elevensy=cmsy10 scaled\magstephalf
\font\ninesy=cmsy9
\font\eightsy=cmsy8
\font\sixsy=cmsy6
\font\fivesy=cmsy5
\skewchar\ninesy='60 \skewchar\eightsy='60 \skewchar\sixsy='60
\skewchar\elevensy'60

\font\eighteenbf=cmbx10 scaled\magstep3

\font\twelvebf=cmbx10 scaled \magstep1
\font\elevenbf=cmbx10 scaled \magstephalf
\font\tenbf=cmbx10
\font\ninebf=cmbx9
\font\eightbf=cmbx8
\font\sixbf=cmbx6
\font\fivebf=cmbx5

\font\elevenit=cmti10 scaled\magstephalf
\font\nineit=cmti9
\font\eightit=cmti8

% Fonts for bold math
\font\eighteenmib=cmmib10 scaled \magstep3
\font\twelvemib=cmmib10 scaled \magstep1
\font\elevenmib=cmmib10 scaled\magstephalf
\font\tenmib=cmmib10
\font\eightmib=cmmib10 scaled 800
\font\sixmib=cmmib10 scaled 600

\font\eighteensyb=cmbsy10 scaled \magstep3
\font\twelvesyb=cmbsy10 scaled \magstep1
\font\elevensyb=cmbsy10 scaled \magstephalf
\font\tensyb=cmbsy10
\font\eightsyb=cmbsy10 scaled 800
\font\sixsyb=cmbsy10 scaled 600

\font\elevenex=cmex10 scaled \magstephalf
\font\tenex=cmex10
\font\eighteenex=cmex10 scaled \magstep3

%%%%%%%%%%%%%%%%%%%%%%%%%%%%
%% Font families

\def\elevenpoint{\def\rm{\fam0\elevenrm}%
  \textfont0=\elevenrm \scriptfont0=\eightrm \scriptscriptfont0=\sixrm
  \textfont1=\eleveni \scriptfont1=\eighti \scriptscriptfont1=\sixi
  \textfont2=\elevensy \scriptfont2=\eightsy \scriptscriptfont2=\sixsy
  \textfont3=\elevenex \scriptfont3=\tenex \scriptscriptfont3=\tenex
  \def\bf{\fam\bffam\elevenbf}%
  \def\it{\fam\itfam\elevenit}%
  \textfont\bffam=\elevenbf \scriptfont\bffam=\eightbf
   \scriptscriptfont\bffam=\sixbf
\normalbaselineskip=13.95pt
  \setbox\strutbox=\hbox{\vrule height9.5pt depth4.4pt width0pt\relax}%
  \normalbaselines\rm}

\elevenpoint %%% default fonts and baselineskip

\def\ninepoint{\def\rm{\fam0\ninerm}%
  \textfont0=\ninerm \scriptfont0=\sixrm \scriptscriptfont0=\fiverm
  \textfont1=\ninei \scriptfont1=\sixi \scriptscriptfont1=\fivei
  \textfont2=\ninesy \scriptfont2=\sixsy \scriptscriptfont2=\fivesy
  \textfont3=\tenex \scriptfont3=\tenex \scriptscriptfont3=\tenex
  \def\it{\fam\itfam\nineit}%
  \textfont\itfam=\nineit
  \def\bf{\fam\bffam\ninebf}%
  \textfont\bffam=\ninebf \scriptfont\bffam=\sixbf
   \scriptscriptfont\bffam=\fivebf
\normalbaselineskip=11pt
  \setbox\strutbox=\hbox{\vrule height8pt depth3pt width0pt\relax}%
  \normalbaselines\rm}

\def\eightpoint{\def\rm{\fam0\eightrm}%
  \textfont0=\eightrm \scriptfont0=\sixrm \scriptscriptfont0=\fiverm
  \textfont1=\eighti \scriptfont1=\sixi \scriptscriptfont1=\fivei
  \textfont2=\eightsy \scriptfont2=\sixsy \scriptscriptfont2=\fivesy
  \textfont3=\tenex \scriptfont3=\tenex \scriptscriptfont3=\tenex
  \def\it{\fam\itfam\eightit}%
  \textfont\itfam=\eightit
  \def\bf{\fam\bffam\eightbf}%
  \textfont\bffam=\eightbf \scriptfont\bffam=\sixbf
   \scriptscriptfont\bffam=\fivebf
\normalbaselineskip=12pt
  \setbox\strutbox=\hbox{\vrule height8.5pt depth3.5pt width0pt\relax}%
  \normalbaselines\rm}

%%%%%%%%%%%%%%%%%%%%%%%%%%%%
%% Font families for bold math in title and section heads

\def\eighteenbold{\def\rm{\fam0\eighteenbf}%
  \textfont0=\eighteenbf \scriptfont0=\twelvebf \scriptscriptfont0=\tenbf
  \textfont1=\eighteenmib \scriptfont1=\twelvemib\scriptscriptfont1=\tenmib
  \textfont2=\eighteensyb \scriptfont2=\twelvesyb\scriptscriptfont2=\tensyb
  \textfont3=\eighteenex \scriptfont3=\tenex \scriptscriptfont3=\tenex
  \def\bf{\fam\bffam\eighteenbf}%
  \textfont\bffam=\eighteenbf \scriptfont\bffam=\twelvebf
   \scriptscriptfont\bffam=\tenbf
\normalbaselineskip=20pt
  \setbox\strutbox=\hbox{\vrule height13.5pt depth6.5pt width0pt\relax}%
\everymath {\fam0 }
\everydisplay {\fam0 }
  \normalbaselines\rm}

\def\elevenbold{\def\rm{\fam0\elevenbf}%
  \textfont0=\elevenbf \scriptfont0=\eightbf \scriptscriptfont0=\sixbf
  \textfont1=\elevenmib \scriptfont1=\eightmib \scriptscriptfont1=\sixmib
  \textfont2=\elevensyb \scriptfont2=\eightsyb \scriptscriptfont2=\sixsyb
  \textfont3=\elevenex \scriptfont3=\elevenex \scriptscriptfont3=\elevenex
  \def\bf{\fam\bffam\elevenbf}%
  \textfont\bffam=\elevenbf \scriptfont\bffam=\eightbf
   \scriptscriptfont\bffam=\sixbf
\normalbaselineskip=14pt
  \setbox\strutbox=\hbox{\vrule height10pt depth4pt width0pt\relax}%
\everymath {\fam0 }
\everydisplay {\fam0 }
  \normalbaselines\bf}

%%%%%%%%%%%%%%%%%%%%%%%%%%%%%%%%%%%%%%%%%%%%%%%%%%%%%%%%%
%% 2) Page set up
\hsize=31pc
\vsize=48pc

\parindent=22pt
\parskip=0pt

\widowpenalty=10000
\clubpenalty=10000

\topskip=12pt

\skip\footins=20pt
\dimen\footins=3in % maximum footnote height

\abovedisplayskip=6.95pt plus3.5pt minus 3pt
\belowdisplayskip=\abovedisplayskip

%% Output routine

\voffset=7pt\hoffset= .7in%7pt magstep1

\newif\iftitle%!

\def\amheadline{\iftitle%
\hbox to\hsize{\hss\currannalsline\hss}\else\line{\ifodd\pageno
\hfill\thetitle\hfill\llap{\elevenrm\folio}\else\rlap{\elevenrm\folio}
\hfill\theauthors\hfill\fi}\fi}

\headline={\amheadline}%!!!
\footline={\global\titlefalse}
%\output={\bindingoffset\plainoutput}

%%%%%%%%%%%%%%%%%%%%%%%%%%%%%%%%%%%%%%%
% 3) Title page

 %#1= Volume number, #2=year of publication
\def\annalsline#1#2{\vfill\eject
\ifodd\pageno\else % first page of article on right.
\line{\hfill}
\vfill\eject\fi
\global\titletrue
\def\currannalsline{\eightrm %Annals of Mathematics,%ANNALS
{\eightbf#1} (#2), \thepages}}

\def\titleheadline#1{\def\one{#1}\ifx\one\empty\else
\def\thetitle{{%\frenchspacing%
\let\\ \relax\eightsc\uppercase{#1}}}\fi}

\newif\ifshort

\let\shorttitle\titleheadline

\def\onpages#1#2{\def\thepages{#1--#2}}

\def\thismuchskip[#1]{\vskip#1pt}
\def\ilook{\ifx\next[ \let\go\thismuchskip\else
\let\go\relax\vskip1pt\fi\go}

\def\institution#1{\def\theinstitutions{\vbox{\baselineskip10pt
\def\and{{\eightrm and }}
\def\\{\futurelet\next\ilook}\eightsc #1}}}
\let\institutions\institution

\newwrite\auxfile

\def\startingpage#1{\def\one{#1}\ifx\one\empty\global\pageno=1\else
\global\pageno=#1\fi
\theoremcount=0 \eqcount=0 \sectioncount=0
\openin1 \jobname.aux \ifeof1
\onpages{#1}{???}
\else\closein1 \relax\input \jobname.aux
\onpages{#1}{\lastpage}
\fi\immediate\openout\auxfile=\jobname.aux
}

\def\endarticle{\ifRefsUsed\global\RefsUsedfalse%
\else\vskip21pt\theinstitutions%
\nobreak\vskip8pt
%\vbox{\thereceived\therevised}%
\fi%
\write\auxfile{\string\def\string\lastpage{\the\pageno}}}

\outer\def\bye{\endarticle\par \vfill \supereject \end}

% variation on code from amsspt.sty ==>
\def\document{\let\fontlist@\relax\let\alloclist@\relax
 \elevenpoint}%%% add for annals!!!

% <=== end of code varied from amsppt.sty

\newif\ifacks
\long\def\acknowledgements#1{\def\one{#1}\ifx\one\empty\else
\vskip-\baselineskip%
\global\ackstrue\footnote{\ \unskip}{*#1}\fi}

\def\title#1{\titleheadline{#1}
\vbox to80pt{\vfill
\baselineskip=18pt
\parindent=0pt
\overfullrule=0pt
\hyphenpenalty=10000
\everypar={\hskip\parfillskip\relax}
\hbadness=10000
\def\\ {\vskip1sp}
\eighteenbold#1\vskip1sp}}

\newif\ifauthor

\def\author#1{\vskip11pt
\hbox to\hsize{\hss\tenrm By \tensc#1\ifacks\global\acksfalse*\fi\hss}
\ifshort\else\xdef\theauthors{{\eightsc\uppercase{#1}}}\fi%
\vskip21pt\global\authortrue\everypar={\global\authorfalse\everypar={}}}

\def\twoauthors#1#2{\vskip11pt
\hbox to\hsize{\hss%
\tenrm By \tensc#1 {\tenrm and} #2\ifacks\global\acksfalse*\fi\hss}
\ifshort\else\xdef\theauthors{{\eightsc\uppercase{#1 and #2}}}\fi%
\vskip21pt
\global\authortrue\everypar={\global\authorfalse\everypar={}}}

%%%%%%%%%%%%%%%%%%%%%%%%%%%%%%%%
%% 4) Section heads, counters

\newcount\theoremcount
\newcount\sectioncount
\newcount\eqcount

\newif\ifspecialnumon

\def\eqnumber=#1 {\global\eqcount=#1 \global\advance\eqcount by-1\relax}
\def\sectionnumber=#1 {\global\sectioncount=#1
\global\advance\sectioncount by-1\relax}
\def\proclaimnumber=#1 {\global\theoremcount=#1
\global\advance\theoremcount by-1\relax}

\newif\ifsection
\newif\ifsubsection

\def\elevenboldmath#1{$#1$\egroup}
\def\mathbold{\hbox\bgroup\elevenbold\elevenboldmath}

\def\section#1{\global\theoremcount=0
\global\eqcount=0
\ifauthor\global\authorfalse\else%
\vskip18pt plus 18pt minus 6pt\fi%
{\parindent=0pt
\everypar={\hskip\parfillskip}%            !!! remove
\def\\ {\vskip1sp}\elevenpoint\bf%
\ifspecialnumon\global\specialnumonfalse$\rm\spnum$%
\gdef\sectnum{$\rm\spnum$}%
\else\interlinepenalty=10000%
\global\advance\sectioncount by1\relax\the\sectioncount%
\gdef\sectnum{\the\sectioncount}%
\fi. \hskip6pt#1%                          !!!add }} and stop here
\vrule width0pt depth12pt}
\hskip\parfillskip%\break%!
\global\sectiontrue%
\everypar={\global\sectionfalse\global\interlinepenalty=0\everypar={}}%
\ignorespaces

}

%%%%%%%%%%%%%%%%%%%%%%%%%%%%%%%%
%% 5) Equation Macros

\newif\ifspequation

\let\eqno\leqno %automatic left side equation numbers %%!!!remove l-eqno

\newif\ifineqalignno
\let\saveleqalignno\leqalignno                        %%!!!remove l-eqno
\def\leqalignno{\let\eqnu\Eeqnu\saveleqalignno}

\let\eqalignno\leqalignno

\def\sectandeqnum{%
\ifspecialnumon\global\specialnumonfalse
$\rm\spnum$\gdef\eqnum{{$\rm\spnum$}}\else\global\firstlettertrue
\global\advance\eqcount by1
\ifappend\applett\else\the\sectioncount\fi.%
\the\eqcount
\xdef\eqnum{\ifappend\applett\else\the\sectioncount\fi.\the\eqcount}\fi}

\def\eqnu{\leqno{\hbox{\elevenrm\ifspequation\else(\fi\sectandeqnum
\ifspequation\global\spequationfalse\else)\fi}}}      %!!! l-eqno

\def\Speqnu{\global\setbox\leqnobox=\hbox{\elevenrm
\ifspequation\else%
(\fi\sectandeqnum\ifspequation\global\spequationfalse\else)\fi}}

\def\Eeqnu{\hbox{\elevenrm
\ifspequation\else%
(\fi\sectandeqnum\ifspequation\global\spequationfalse\else)\fi}}

\newif\iffirstletter
\global\firstlettertrue
\def\eqletter#1{\global\specialnumontrue\iffirstletter\global\firstletterfalse
\global\advance\eqcount by1\fi
\gdef\spnum{\the\sectioncount.\the\eqcount#1}\eqnu}

%%% Split math
\newbox\leqnobox
\def\outsideeqnu#1{\global\setbox\leqnobox=\hbox{#1}}

\def\eatone#1{}

%% Vertically centers equation number.
\def\dosplit#1#2{\vskip-.5\abovedisplayskip
\setbox0=\hbox{$\let\eqno\outsideeqnu%
\let\eqnu\Speqnu\let\leqno\outsideeqnu#2$}%
\setbox1\vbox{\noindent\hskip\wd\leqnobox\ifdim\wd\leqnobox>0pt\hskip1em\fi%
$\displaystyle#1\mathstrut$\hskip0pt plus1fill\relax
\vskip1pt
\line{\hfill$\let\eqnu\eatone\let\leqno\eatone%
\displaystyle#2\mathstrut$\ifmathqed~~\qed\fi}}%
\copy1
\ifvoid\leqnobox
\else\dimen0=\ht1 \advance\dimen0 by\dp1
\vskip-\dimen0
\vbox to\dimen0{\vfill
\hbox{\unhbox\leqnobox}
\vfill}
\fi}

\everydisplay{\lookforbreak}

\long\def\lookforbreak #1$${\def\mathone{#1}
\expandafter\testforbreak\mathone\splitmath @}

\def\testforbreak#1\splitmath #2@{\def\mathtwo{#2}\ifx\mathtwo\empty%
#1$$%
\ifmathqed\vskip-\belowdisplayskip
\setbox0=\vbox{\let\eqno\relax\let\eqnu\relax$\displaystyle#1$}%
\vskip-\ht0\vskip-3.5pt\hbox to\hsize{\hfill\qed}
\vskip\ht0\vskip3.5pt\fi
\else$$\vskip-\belowdisplayskip
\vbox{\dosplit{#1}{\let\eqno\eatone
\let\splitmath\relax#2}}%
\nobreak\vskip.5\belowdisplayskip
\noindent\ignorespaces\fi}

%% Proof box to be used when proof ends with equation.

\newif\ifmathqed

%%%%%%%%%%%%%%%%%%%%%%%%%%%%%
%% \mtable, Math table to make binary table easily

%% Use:
% \mtable
% &n_1&n_2&n_3&n_4&n_5&n_6\cr
% \Delta_1&M_3&M_2&0&0&0&0\cr
% \Delta_2&0&0&M_1&M_3&0&0\cr
% \endmtable

\newcount\linenum
\newcount\colnum

%++
\def\spline{\omit&\multispan{\the\colnum}{\hrulefill}\cr}
\def\colcounter{\ifnum\linenum=1\global\advance\colnum by1\fi}

\def\everyline{\noalign{\global\advance\linenum by1\relax}%
\ifnum\linenum=2\spline\fi}

\def\mtable{\bgroup\offinterlineskip
\everycr={\everyline}\global\linenum=0
\halign\bgroup\vrule height 10pt depth 4pt width0pt
\hfill$##$\hfill\hskip6pt\ifnum\linenum>1
\vrule\fi&&\colcounter\hskip12pt\hfill$##$\hfill\hskip12pt\cr}

\def\endmtable{\crcr\egroup\egroup}

%%%%%%%%%%%%%%%%%%%%%%%%%%%%%
% Array

%% Will work in math or in text, will be in math mode inside array.
%% For each column desired supply
%% r, l, or c, for right, left, or center orientation of that column.
%% End each line with \\.

%% To use:
%  \array ccc*
%  x_s\leq a_1\\
%  a_s<x_s^s<b_s\\
%  x_s\geq a_1
%  \endarray

\def\xast{*}
\newcount\intable
\newcount\mathcol
\newcount\savemathcol
\newcount\topmathcol
\newdimen\arrayhspace
\newdimen\arrayvspace

\arrayhspace=8pt % horizontal space between columns, (half this width
                 %  will horizontally precede and follow the array)
\arrayvspace=12pt % vertical space between lines

\newif\ifdollaron

\def\mathalign#1{\def\arg{#1}\ifx\arg\xast%
\let\go\relax\else\let\go\mathalign%
\global\advance\mathcol by1 %
\global\advance\topmathcol by1 %
\expandafter\def\csname  mathcol\the\mathcol\endcsname{#1}%
\fi\go}

\def\arraypickapart#1]#2*{\if#1c \ifmmode\vcenter\else
\global\dollarontrue$\vcenter\fi\else%
\if#1t\vtop\else\if#1b\vbox\fi\fi\fi\bgroup%
\def\one{#2}}

\def\arraystrut{\vrule height .7\arrayvspace depth .3\arrayvspace width 0pt}

\def\array#1#2*{\def\firstarg{#1}%
\if\firstarg[ \def\two{#2} \expandafter\arraypickapart\two*\else%
\ifmmode\vcenter\else\vbox\fi\bgroup \def\one{#1#2}\fi%
\global\everycr={\noalign{\global\mathcol=\savemathcol\relax}}%
\def\\ {\cr}%
\global\advance\intable by1 %
\ifnum\intable=1 \global\mathcol=0 \savemathcol=0 %
\else \global\advance\mathcol by1 \savemathcol=\mathcol\fi%
\expandafter\mathalign\one*%
\mathcol=\savemathcol %
\halign\bgroup&\hskip.5\arrayhspace\arraystrut%
\global\advance\mathcol by1 \relax%
\expandafter\if\csname mathcol\the\mathcol\endcsname r\hfill\else%
\expandafter\if\csname mathcol\the\mathcol\endcsname c\hfill\fi\fi%
$\displaystyle##$%
\expandafter\if\csname mathcol\the\mathcol\endcsname r\else\hfill\fi\relax%
\hskip.5\arrayhspace\cr}

\def\endarray{\crcr\egroup\egroup%
\global\mathcol=\savemathcol %
\global\advance\intable by -1\relax%
\ifnum\intable=0 %
\ifdollaron\global\dollaronfalse $\fi
\loop\ifnum\topmathcol>0 %
\expandafter\def\csname  mathcol\the\topmathcol\endcsname{}%
\global\advance\topmathcol by-1 \repeat%
\global\everycr={}\fi%
}

\def\big#1{{\hbox{$\left#1\vbox to 10pt{}\right.\n@space$}}}
\def\Big#1{{\hbox{$\left#1\vbox to 13pt{}\right.\n@space$}}}
\def\bigg#1{{\hbox{$\left#1\vbox to 16pt{}\right.\n@space$}}}
\def\Bigg#1{{\hbox{$\left#1\vbox to 19pt{}\right.\n@space$}}}

%%%%%%%%%%%%%%%%%%%%%%%%%%%%%%%%%%%%%%%%%%%%%%%%%%%%%%%%%%%%%%%%
% 6) Figure and Table Captions.

\def\figcaption#1#2#3{\topinsert
\vskip4pt %<===topadjust to match height of ascenders on opposing page.
\vbox to#3{\vfill}\vskip1sp
\setbox0=\hbox{\eightsc Figure #1.\hskip12pt\eightpoint #2}
\ifdim\wd0>\hsize
\noindent\eightsc Figure #1.\hskip12pt\eightpoint #2
\else
\centerline{\eightsc Figure #1.\hskip12pt\eightpoint #2}
\fi
\vskip16pt
\endinsert}

\def\wfig#1#2#3{\topinsert
\vskip4pt %<===topadjust to match height of ascenders on opposing page.
\hbox to\hsize{\hss\vbox{\hrule height .25pt width #3
\hbox to #3{\vrule width .25pt height #2\hfill\vrule width .25pt height #2}
\hrule height.25pt}\hss}
\vskip1sp
\centerline{\eightsc Figure #1}
\vskip16pt
\endinsert}

\def\wfigcaption#1#2#3#4{\topinsert
\vskip4pt %<===topadjust to match height of ascenders on opposing page.
\hbox to\hsize{\hss\vbox{\hrule height .25pt width #4
\hbox to #4{\vrule width .25pt height #3\hfill\vrule width .25pt height #3}
\hrule height.25pt}\hss}
\vskip1sp
\setbox0=\hbox{\eightsc Figure #1.\hskip12pt\eightpoint\rm #2}
\ifdim\wd0>\hsize
\noindent\eightsc Figure #1.\hskip12pt\eightpoint\rm #2\else
\centerline{\eightsc Figure #1.\hskip12pt\eightpoint\rm #2}\fi
\vskip16pt
\endinsert}

\def\tabcaption#1#2{\vskip6pt
\setbox0=\hbox{\eightsc Table #1.\hskip12pt\eightpoint #2}
\ifdim\wd0>\hsize
\noindent\eightsc Table #1.\hskip12pt\eightpoint #2
\else
\centerline{\eightsc Table #1.\hskip12pt\eightpoint #2}
\fi
\vskip6pt}

\def\endinsert{\egroup\if@mid\dimen@\ht\z@\advance\dimen@\dp\z@
\advance\dimen@ 12\p@\advance\dimen@\pagetotal\ifdim\dimen@ >\pagegoal
\@midfalse\p@gefalse\fi\fi\if@mid\smallskip\box\z@\bigbreak\else
\insert\topins{\penalty 100 \splittopskip\z@skip\splitmaxdepth\maxdimen
\floatingpenalty\z@\ifp@ge\dimen@\dp\z@\vbox to\vsize {\unvbox \z@
\kern -\dimen@ }\else\box\z@\nobreak\smallskip\fi}\fi\endgroup}

\def\pagecontents{
\ifvoid\topins \else\iftitle\else
\unvbox \topins \fi\fi \dimen@ =\dp \@cclv \unvbox
\@cclv
\ifvoid\topins\else\iftitle\unvbox\topins\fi\fi
\ifvoid \footins \else \vskip \skip \footins \footnoterule
\unvbox \footins \fi \ifr@ggedbottom \kern -\dimen@ \vfil \fi}

%%%%%%%%%%%%%%%%%%%%%%%%%%%%%%%%%%%%%%%%%%%%%%%%%%%%%%%%%%%%%%%%
% 7) End Matter

\newif\ifappend

\def\appendix#1#2{\def\applett{#1}\def\two{#2}%
\global\appendtrue
\global\theoremcount=0
\global\eqcount=0
\vskip18pt plus 18pt
\vbox{\parindent=0pt
\everypar={\hskip\parfillskip}
\def\\ {\vskip1sp}\elevenbold Appendix%
\ifx\applett\empty\gdef\applett{A}\ifx\two\empty\else.\fi%
\else\ #1.\fi\hskip6pt#2\vskip12pt}%
\global\sectiontrue%
\everypar={\global\sectionfalse\everypar={}}\nobreak\ignorespaces}

\newif\ifRefsUsed
\long\def\references{\global\RefsUsedtrue\vskip21pt
\theinstitutions
\global\everypar={}\global\bibnum=0
\vskip20pt\goodbreak\bgroup
\vbox{\centerline{\eightsc References}\vskip6pt}%
\ifdim\maxbibwidth>0pt
\leftskip=\maxbibwidth%
\parindent=-\maxbibwidth%
\else
\leftskip=18pt%
\parindent=-18pt%
\fi
\ninepoint
\frenchspacing
\nobreak\ignorespaces\everypar={\amref}%
}

\def\endreferences{\vskip1sp\egroup\global\everypar={}%
\nobreak\vskip8pt\vbox{\thereceived\therevised}
}

\newcount\bibnum

\def\amref#1 {\global\advance\bibnum by1%
\immediate\write\auxfile{\string\expandafter\string\def\string\csname
\space #1croref\string\endcsname{[\the\bibnum]}}%
\leavevmode\hbox to18pt{\hbox to13.2pt{\hss[\the\bibnum]}\hfill}}

\def\bibline{\hbox to30pt{\hrulefill}\/\/}

\def\name#1{{\eightsc#1}}

\newdimen\maxbibwidth
\def\AuthorRefNames [#1] {%
\immediate\write\auxfile{\string\def\string\cite\string##1{[\string##1]}}

\def\amref{\spamref}
\setbox0=\hbox{[#1] }\global\maxbibwidth=\wd0\relax}

\def\spamref[#1] {\leavevmode\hbox to\maxbibwidth{\hss[#1]\hfill}}

%%%%%%%%%%%%%%%%%%%%%%%%%%%%%%%%%%%%%%%%%%%%%%%%%%%%%%%%%%%%%%%%
%% 8) Footnotes

\def\footnoterule{\kern-3pt\hrule width1in height.5pt\kern2.5pt}

\def\footnote#1#2{%
\plainfootnote{#1}{{\eightpoint\normalbaselineskip11pt
\normalbaselines#2}}}

\def\vfootnote#1{%
\insert \footins \bgroup \eightpoint\baselineskip11pt
\interlinepenalty \interfootnotelinepenalty
\splittopskip \ht \strutbox \splitmaxdepth \dp \strutbox \floatingpenalty
\@MM \leftskip \z@skip \rightskip \z@skip \spaceskip \z@skip
\xspaceskip \z@skip
{#1}$\,$\footstrut \futurelet \next \fo@t}

%%%%%%%%%%%%%%%%%%%%%%%%%%%%%%%%%%%%%%%%%%%%%%%%%%%%%%%%%%%%%%%%
%% 9) Theorem type environments

\newif\iffirstadded
\newif\ifadded

\def\addedlett{}

\def\alltheoremnums{%
\ifspecialnumon\global\specialnumonfalse
\ifadded\global\addedfalse
\iffirstadded\global\firstaddedfalse
\global\advance\theoremcount by1 \fi
\ifappend\applett\else\the\sectioncount\fi.\the\theoremcount\addedlett%
\xdef\theoremnum{\ifappend\applett\else\the\sectioncount\fi.%
\the\theoremcount\addedlett}%
\else$\rm\spnum$\def\theoremnum{{$\rm\spnum$}}\fi%
\else\global\firstaddedtrue
\global\advance\theoremcount by1
\ifappend\applett\else\the\sectioncount\fi.\the\theoremcount%
\xdef\theoremnum{\ifappend\applett\else\the\sectioncount\fi.%
\the\theoremcount}\fi}

\def\allcorolnums{%
\ifspecialnumon\global\specialnumonfalse
\ifadded\global\addedfalse
\iffirstadded\global\firstaddedfalse
\global\advance\corolcount by1 \fi
\the\corolcount\addedlett%
\else$\rm\spnum$\def\corolnum{$\rm\spnum$}\fi%
\else\global\advance\corolcount by1
\the\corolcount\fi}

%% use for Theorem, Corollary, Lemma, Proposition, Demonstration and similar.

\newcount\corolcount
\def\xcorol{Corollary}
\def\xtheorem{Theorem}
\def\xmaintheorem{Main Theorem}

\newif\ifthtitle

\let\saverparen)
\let\savelparen(
\def\rmparenl{{\rm(}}
\def\rmparenr{{\rm\/)}}
{
\catcode`(=13
\catcode`)=13
\gdef\makeparensRM{\catcode`(=13\catcode`)=13\let(=\rmparenl%
\let)=\rmparenr%
\everymath{\let(\savelparen%
\let)\saverparen}%
\everydisplay{\let(\savelparen%
\let)\saverparen\lookforbreak}}}

\medskipamount=8pt plus.1\baselineskip minus.05\baselineskip

\def\rmtext#1{\hbox{\rm#1}}

\def\proclaim#1{\vskip-\lastskip
\def\one{#1}\ifx\one\xtheorem\global\corolcount=0\fi
\ifsection\global\sectionfalse\vskip-6pt\fi
\medskip
{\elevensc#1}%
\ifx\one\xmaintheorem\global\corolcount=0
\gdef\theoremnum{Main Theorem}\else%
\ifx\one\xcorol\ \allcorolnums\else\ \alltheoremnums\fi\fi%
\ifthtitle\ \global\thtitlefalse{\rm(\thethtitle)}\fi.%
\hskip1em\bgroup\let\text\rmtext\makeparensRM\it\ignorespaces}

\def\nonumproclaim#1{\vskip-\lastskip
\def\one{#1}\ifx\one\xtheorem\global\corolcount=0\fi
\ifsection\global\sectionfalse\vskip-6pt\fi
\medskip
{\elevensc#1}.\ifx\one\xmaintheorem\global\corolcount=0
\gdef\theoremnum{Main Theorem}\fi\hskip.5pc%
\bgroup\it\makeparensRM\ignorespaces}

\def\endproclaim{\egroup\medskip}

%% Use demo for Proof, Proof of, Definition, Example,
%% Remark, Case, Subcase, Conjecture, Note, Notation,
%% Convention, Construction and Step.
%% Any other use for demo will format similar to `Proof.'

\def\xproof{Proof}
\def\xremark{Remark}
\def\xcase{Case}
\def\xsubcase{Subcase}
\def\xconjecture{Conjecture}
\def\xstep{Step}
\def\xof{of}

\def\deconstruct#1 #2 #3 #4 #5 @{\def\one{#1}\def\two{#2}\def\three{#3}%
\def\four{#4}%
\ifx\two\empty #1\else%
\ifx\one\xproof%
\ifx\two\xof%
  \ifx\three\xcorol Proof of Corollary \rm#4\else%
     \ifx\three\xtheorem Proof of Theorem \rm#4\else\xone\fi%
  \fi\fi%
\else\xone\fi\fi.}

\def\pickup#1 {\def\this{#1}%
\ifx\this\xproof\global\let\go\demoproof
\global\let\enddemo\endproof\else
\ifx\this\xremark\global\let\go\demoremark\else
\ifx\this\xcase\global\let\go\demostep\else
\ifx\this\xsubcase\global\let\go\demostep\else
\ifx\this\xconjecture\global\let\go\demostep\else
\ifx\this\xstep\global\let\go\demostep\else
\global\let\go\demoproof\fi\fi\fi\fi\fi\fi}

\newif\ifnonum
\def\demo#1{\vskip-\lastskip
\ifsection\global\sectionfalse\vskip-6pt\fi
\def\one{#1 }\def\two{#1*}%
\setbox0=\hbox{\expandafter\pickup\one}\expandafter\go\two}

\def\numbereddemo#1{\vskip-\lastskip
\ifsection\global\sectionfalse\vskip-6pt\fi
\def\two{#1*}%
\expandafter\demoremark\two}

\def\demoproof#1*{\medskip\def\xone{#1}
{\ignorespaces\it\expandafter\deconstruct\xone {} {} {} {} {} @%
\unskip\hskip6pt}\rm\ignorespaces}

\def\demoremark#1*{\medskip
{\it\ignorespaces#1\/} \ifnonum\global\nonumtrue\else
 \alltheoremnums\unskip.\fi\hskip1pc\rm\ignorespaces}

\def\demostep#1 #2*{\vskip4pt
{\it\ignorespaces#1\/} #2.\hskip1pc\rm\ignorespaces}

\def\enddemo{\medskip}

\def\endproof{\ifmathqed\global\mathqedfalse\medskip\else
\parfillskip=0pt~~\hfill\qed\medskip
\fi\global\parfillskip0pt plus 1fil\relax
\gdef\enddemo{\medskip}}

\def\qed{\vbox{\hrule\hbox{\vrule height6pt\hskip6pt\vrule}\hrule}}

%% Proof box to be used in a \proclaim{}...\endproclaim environment

\def\proofbox{\parfillskip=0pt~~\hfill\qed\vskip1sp\parfillskip=
0pt plus 1fil\relax}

%%%%

%%%%%%%%%%%%%%%%%%%%%
%% 10) CrossRefs

%%% Generic crossreferencing
%%% to use: \label\nameoflabel* (will give the page number when referenced)

% Commands to access current state of counter, for cross-referencing
% \sectnum
% \theoremnum
% \eqnum

%%% You can make another definition that includes counters and/or the
%%% page number and access this information as the second argument:
%%% \label\yourlabelname[2.13]*

%%% Since this method of cross-referencing relies
%%% on an auxiliary file, the first time you tex the file
%%% you will get `??' when you write \ref\nameoflabel.
%%% When you TeX the file the second time the auxiliary file
%%% will be input and \ref\nameoflabel will produce the cross-ref.

\def\stripbs#1#2*{\def\one{#2}}

\def\emptyspace{ }
\def\nextthing{}
\def\newline{***}
\def\eatone#1{ }

\def\lookatspace#1{\ifcat\noexpand#1\ \else%
\gdef\nextthing{}\xdef\next{#1}%
\ifx\next\emptyspace%
\let\nextthing\emptyspace\else\ifx\next\newline%
\gdef\nextthing{\eatone}\fi\fi\fi\egroup\nextthing#1}

{\catcode`\^^M=\active%
\gdef\spacer{\bgroup\catcode`\^^M=\active%
\let^^M=\newline\obeyspaces\lookatspace}}

\def\ref#1{\seeifdefined{#1}\expandafter\csname\one\endcsname\spacer}

\def\cite#1{\expandafter\ifx\csname#1croref\endcsname\relax[??]\else
\csname#1croref\endcsname\fi\spacer}

%% for testing in \label and \ref to see if term already labeled.

\def\seeifdefined#1{\expandafter\stripbs\string#1croref*%
\crorefdefining{#1}}

\newif\ifcromessage
\global\cromessagetrue

\def\crorefdefining#1{\ifdefined{\one}{}
{\ifcromessage\global\cromessagefalse%
\message{\spaces\spaces\spaces\spaces\spaces\spaces\spaces}%
\message{<Undefined reference.}%
\message{Please TeX file once more to have accurate cross-references.>}%
\message{\spaces\spaces\spaces\spaces\spaces\spaces\spaces}\fi[??]}}

\def\label#1#2*{\gdef\ctest{#2}%
\xdef\currlabel{\string#1croref}
\expandafter\seeifdefined{#1}%
\ifx\empty\ctest%
\xdef\labelnow{\write\auxfile{\noexpand\def\currlabel{\the\pageno}}}%
\else\xdef\labelnow{\write\auxfile{\noexpand\def\currlabel{#2}}}\fi%
\labelnow}

\def\ifdefined#1#2#3{\expandafter\ifx\csname#1\endcsname\relax%
#3\else#2\fi}

%%%%%%%%%%%%%%%%%%%%%
%% 11) Listing

%% To use with asterisks:

%%%%%%%%%%%%%%%%%%%%%%
%% 12) Article and Journal Table of Contents

\def\articlecontents{
\vskip20pt\centerline{\bf Table of Contents}\everypar={}\vskip6pt
\bgroup \leftskip=3pc \parindent=-2pc
\def\item##1{\vskip1sp\indent\hbox to2pc{##1.\hfill}}}

\def\endcontents{\vskip1sp\leftskip=0pt\egroup}

\def\journalcontents{\vfill\eject
\def\currannalsline{\hfill}
\global\titletrue
\vglue3.5pc
\centerline{\tensc\hskip12pt TABLE OF CONTENTS}\everypar={}\vskip30pt
\bgroup \leftskip=34pt \rightskip=-12pt \parindent=-22pt
  \def\\ {\vskip1sp\noindent}
\def\pagenum##1{\unskip\parfillskip=0pt\dotfill##1\vskip1sp
\parfillskip=0pt plus 1fil\relax}
\def\name##1{{\tensc##1}}}

%% default values

\institution{}
\onpages{0}{0}
\def\lastpage{???}
\def\thetitle{Title ???}
\def\theauthors{Authors ???}
\def\thereceived{}
\def\therevised{}

\gdef\split{\relaxnext@\ifinany@\let\next\insplit@\else
 \ifmmode\ifinner\def\next{\onlydmatherr@\split}\else
 \let\next\outsplit@\fi\else
 \def\next{\onlydmatherr@\split}\fi\fi\let\eqnu\xspliteqnu\next}

\gdef\align{\relaxnext@\ifingather@\let\next\galign@\else
 \ifmmode\ifinner\def\next{\onlydmatherr@\align}\else
 \let\next\align@\fi\else
 \def\next{\onlydmatherr@\align}\fi\fi\let\eqnu\xspliteqnu\next}

\def\spliteqnu{{\tenrm\sectandeqnum}\relax}

\def\xspliteqnu{\tag\spliteqnu}

\catcode`@=12

\document

%-------------- Publisher's entries --------------------

\annalsline{}{}
\startingpage{1}     %numeration
%%\received{??}
%%\revised{??}

\comment
\nopagenumbers
\headline{\ifnum\pageno=1\hfil\else \rightheadline\fi}
%{\ifodd\pageno\rightheadline \else \leftheadline\fi}\fi}%%after else
\def\rightheadline{\hfil\eightit
%! Running title (odd page)
The Macdonald conjecture
\quad\eightrm\folio}

\voffset=2\baselineskip
\endcomment

%\magnification=\magstep1

%--------------- Author macros ---------------
%                   MACROS
%
%                                 AUX
%
%

\def\cite#1{[#1]}

\def\lastpage{16}

%
%                      endaux
%

\def\iif{\quad\hbox{ if }\quad}

\def\for{\  \hbox{ for } \ }
\def\if{ \ \hbox{ if } \ }
\def\when{ \ \hbox{ when } \ }
\def\where{\  \hbox{ where } \ }
\def\and{\  \hbox{ and } \ }

\def\equal{\buildrel  def \over =}

\def\la{\lambda}
\def\La{\Lambda}
\def\om{\omega}

\def\th{\theta}
\def\al{\alpha}
\def\be{\beta}
\def\ga{\gamma}
\def\ep{\epsilon}

\def\de{\delta}
\def\De{\Delta}

\def\si{\sigma}

\def\Ga{\Gamma}
\def\ze{\zeta}

\def\bio{b^\iota}
\def\aio{a^\iota}

\def\gio{\g^\iota}
\def\Bio{B^\iota}

    %from copy, ell

\def\vep{\varepsilon}

\def\tal{\tilde{\alpha}}
\def\tbe{\tilde{\beta}}

\def\tw{\tilde w}

\def\tz{\tilde z}
\def\tb{\tilde b}

\def\hH{\hat{H}}

\def\hY{\hat{Y}}

\def\hT{\hat{T}}

\def\hw{\hat{w}}

\def\hv{\hat{v}}

\def\C{\bold{C}}
\def\R{\bold{R}}

\def\Z{\bold{Z}}

\def\one{\bold{1}}

\def\0{\bold{0}}

\def\C{\hbox{\bf C}}

%\def\bs{\hbox{\bf S}}          %ell
% macdonald

\def\f{\Cal{F}}
\def\t{\Cal{T}}

\def\l{\Cal{L}}

\def\y{\Cal{Y}}

\def\x{\Cal{X}}
\def\g{\Cal{G}}

\def\w{\Cal{W}}

\font\germ=eufb10 at 12pt
%\font\germm=eufb9 at 12pt
%\font\germ=eufm9 at 12pt
\def\goth#1{\hbox{\germ #1}}

\def\TT{\goth{T}}
\def\HH{\goth{H}}
\def\FF{\goth{F}}

\font\smm=msbm10 at 12pt %msym in charlie
\def\symbol#1{\hbox{\smm #1}}
\def\lsmash{{\symbol n}}

%endmacros

%------------------------------------------------------------------
%-------------- Author entries --------------------

%\comment                               %to remove the title
\title
{Difference-elliptic operators\\
and root systems
}
 %Article title
\shorttitle{ Difference-elliptic operators}
 % Shortened version for headline title

% Acknowledgements: Please enter all acknowledgements here.
\acknowledgements{
Partially supported by NSF grant DMS--9301114, C.N.R.S.,  and UNC Research
Counsel.
}

% Please uncomment and use appropriate command:
\author{ Ivan Cherednik}
%\twoauthors{}{}
%\authors{}% Separate each author with a comma and a space.

% Institution:
%% If more than one institution represented, please separate
%% with \\ , i.e.,
%% \institutions{University of Illinois at Chicago, Chicago, IL\\
%% Cornell University, Ithaca, NY}

\institutions{
University of North Carolina at Chapel Hill,
Chapel Hill, N.C. 27599-3250
\\ Internet: chered\@math.unc.edu
}

%\endcomment                              %to remove the title
%-------------- Article Text--------------------

%\intro %(Optional, Introduction)
%
%
%
%                        INTRO
%
%
%{\bf 0. Introduction.}
\vfil
Recently a new technique in the harmonic analysis on
symmetric spaces was suggested based on certain
remarkable representations of affine and
double affine Hecke algebras in terms of Dunkl and
Demazure operators instead of Lie groups and Lie algebras.
In the classic case (see
[O,H,C5]) it resulted  (among  other
applications) in a new  theory of radial
part of Laplace operators and their deformations
including a related concept of the Fourier transform
(see [DJO]). Some observations indicate that there can
 be connections with
 the so-called $W_\infty$-algebras.

In papers [C1,C2,C4] the analogous difference methods
 were developed   to generalize
the operators constructed by Macdonald
(corresponding to the minuscule and certain similar weights)
and  those considered in
[N] and other works on $q$-symmetric spaces.
It is quite likely that the Fourier transform is self-dual
in the difference setting
(in contrast to the classical theory).

Paper [C3] is devoted to the differential-elliptic case
presumably corresponding to the Kac-Moody algebras.
Presumably because the ways of extending
the traditional harmonic analysis to these algebras are
still rather obscure although there are interesting projects.
 In the present  paper we demonstrate that the
new technique works well even in the most general
difference-elliptic case
conjecturally  corresponding to the $q$-Kac-Moody
algebras considered at the critical level.

We discuss here only the construction of the generalized
radial (zonal)   Laplace operators.
These operators are closely related to
the so-called quantum many-body problem
(Calogero,  Sutherland, Moser, Olshanetsky,
 Perelomov),
the
con\-for\-mal field
 theory  (Knizhnik- Zamolodchikov equations),
and  the classic theory of the
hypergeometric functions. They are expected to have
applications to the characters of the Kac-Moody algebras
and in Arithmetic.
The natural problem is to extend
 the  Macdonald theory [M1,M2,C2] to the elliptic case.

We also
connect our operators
with the difference-elliptic Ruijsenaars operators
from [R] (corresponding to the minuscule weights of type $A$)
generalizing in its turn the Olshanetsky-Perelomov differential
elliptic operators.

At the end of the paper we show that the
 monodromy
of the trigonometric KZ equation from [C4] leads
(when properly defined) to a new
invariant R-matrix of elliptic type (which might be
of a certain importance to establish the relation
between double affine Hecke algebras and proper "elliptic"
quantum groups).
 The invariance means that the monodromy elements
for $W$-conjugated simple reflections are conjugated
with respect to the action of the Weyl group $W$.
We note that the monodromy elements (matrices) always
satisfy the braid relations. However if one defines the
monodromy representation following
[FR] and other papers devoted to
trigonometric difference KZ equations then it is
never  invariant. The monodromy matrices
for simple reflections  are  connected  in a more
complicated way (the relations are similar to the
 so-called Star-Triangle identities).

This work was started at
 the Laboratoire de Mathematiques
Fondamentales (Universit\'e Paris 6).
 I'd like to thank
 A.Connes, A.Joseph, and  R. Rentschler
for the kind invitation
and hospitality. I am grateful to P.Etingof
 for a useful discussion.

%
%
%		Section 1
%
%
%\vskip 10pt
\section { Affine root systems}
Let $R=\{\al\}   \subset \R^n$ be a root system of type $A,B,...,F,G$
with respect to a euclidean form $(z,z')$ on $\R^n \ni z,z'$.
The latter is unique up to proportionality.
We assume that $(\al,\al)=2$ for long $\al$.
Let us  fix the set $R_{+}$ of positive  roots ($R_-=-R_+$),
the corresponding simple
roots $\al_1,...,\al_n$, and  their dual counterparts
$a_1 ,..., a_n,  a_i =\al_i^\vee, \where \al^\vee =2\al/(\al,\al)$.
The dual fundamental weights
$b_1,...,b_n$  are determined from the relations  $ (b_i,a_j)=
\de_i^j $ for the
Kronecker delta. We will also introduce the lattices
$$
\eqalignno{
& A=\oplus^n_{i=1}\Z a_i \subset B=\oplus^n_{i=1}\Z b_i,
}
$$
and  $A_\pm, B_\pm$  for $\Z_{\pm}=\{m\in\Z, \pm m\ge 0\}$
instead of $\Z$. (In the standard notations, $A= Q^\vee,\
B = P^\vee $.) Later on,
$$
\eqalign{
&\nu_{\al}\ =\ (\al,\al),\  \nu_i\ =\ \nu_{\al_i}, \
\nu_R\ = \{\nu_{\al}, \al\in R\}, \cr
&\rho_\nu\ =\ (1/2)\sum_{\nu_{\al}=\nu} \al \ =
\ (\nu/2)\sum_{\nu_i=\nu}  b_i, \for\al\in R_+.}
%\eqno(0.1)
\eqnu
$$

The vectors $\ \tal=[\al,k] \in
\R^n\times \R \subset \R^{n+1}$
for $\al \in R, k \in \Z $
form the {\it affine root system}
$R^a \supset R$ ( $z\in \R^n$ are identified with $ [z,0]$).
We add  $\al_0 \equal [-\th,1]$ to the  simple roots
for the {\it maximal root} $\th \in R$.
The corresponding set $R^a_+$ of positive roots coincides
with $R_+\cup \{[\al,k],\  \al\in R, \  k > 0\}$.

We denote the Dynkin diagram and its affine completion with
$\{\al_j,0 \le j \le n\}$ as the vertices by $\Ga$ and $\Ga^a$.
Let $m_{ij}=2,3,4,6$\  if $\al_i\and\al_j$ are joined by 0,1,2,3 laces
respectively.
The set of
the indices of the images of $\al_0$ by all
the automorphisms of $\Ga^a$ will be denoted by $O$ ($O=\{0\}
\for E_8,F_4,G_2$). Let $O^*={r\in O, r\neq 0}$.
The elements $b_r$ for $r\in O^*$ are the so-called minuscule
weights ($(b_r,\al)\le 1$ for
$\al \in R_+$).

%Without going into detail, we mention that $(\th^\vee,\al) \le 1$ for
%$\th \neq \al \in R_+$. Moreover, $\th^\vee=\sum_i b_i,
%\where m_{i0}>2$.
%The multiplicity   is also not more than 1 for $r\in O^*$ (i.e. $b_r$
%are minuscule co-weights). For instance,
%$(b_r,\th)=1$  (see [B,V,C4]).

Given $\tal=[\al,k]\in R^a,  \ b \in B$, let
$$
\eqalignno{
&s_{\tal}(\tz)\ =\  \tz-(z,\al^\vee)\tal,\
\ b'(\tz)\ =\ [z,\ze-(z,b)]
&\eqnu
%&(1.1)
}
$$
for $\tz=[z,\ze] \in \R^{n+1}$.

The {\it affine Weyl group} $W^a$ is generated by all $s_{\tal}$
(we write $W^a = <s_{\tal}, \tal\in R_+^a>)$. One can take
the simple reflections $s_j=s_{\al_j}, 0 \le j \le n,$ as its
generators and introduce the corresponding notion of the
length. This group is
the semi-direct product $W\lsmash A'$ of
its subgroups $W=<s_\al,
\al \in R_+>$ and $A'=\{a', a\in A\}$, where
$$
\eqalignno{
& a'=\ s_{\al}s_{[\al,1]}=\ s_{[-\al,1]}s_{\al}\for a=\al^{\vee},
\ \al\in R.
&\eqnu
%&(1.2)
}
$$

The {\it extended Weyl group} $ W^b$ generated by $W\and B'$
(instead of $A'$) is isomorphic to $W\lsmash B'$:
$$
\eqalignno{
&(wb')([z,\ze])\ =\ [w(z),\ze-(z,b)] \for w\in W, b\in B.
%&(1.3)
&\eqnu
}
$$

\proclaim{Definition }%1.1 }
{i) } Given $b_+\in B_+$, let
$$
\eqalignno{
&\om_{b_+} = w_0w^+_0  \in  W,\ \pi_{b_+} =
b'_+(\om_{b_+})^{-1}
\ \in \ W^b, \ \om_i=\om_{b_i},\pi_i=\pi_{b_i},
&\eqnu
%&(1.4)
}
$$
where $w_0$ (respectively, $w^+_0$) is the longest element in $W$
(respectively, in $ W_{b_+}$ generated by $s_i$ preserving $b_+$)
relative to the
set of generators $\{s_i\}$ for $i >0$.

{ii)}
If $b$ is arbitrary then there exist  unique elements $w\in W,\
b_+ \in B_+$
such that  $b=w(b_+)$ and
$(\al,b_+) \neq 0 \if (-\al) \in R_+ \ni w(\al)$. We set
$$
\eqalignno{
&b'=\pi_b\om_b,\where \om_b = \om_{b_+}w^{-1},
\pi_b = w\pi_{b_+}.
%&(1.5)
&\eqnu
}
$$
\endproclaim
\proofbox

We will mostly use  the
elements $\pi_r=\pi_{b_r}, r \in O$. They leave $\Ga^a$ invariant
and form a group denoted by $\Pi$,
 which is isomorphic to $B/A$ by the natural
projection $\{b_r \to \pi_r\}$. As to $\{\om_r\}$,
they preserve the set $\{-\th,\al_i, i>0\}$.
The relations $\pi_r(\al_0)= \al_r= (\om_r)^{-1}(-\th)
$ distinguish the
indices $r \in O^*$. These elements are important because
due to [B,V,C4]:
$$
\eqalignno{
& W^b  = \Pi \lsmash W^a, \where
  \pi_rs_i\pi_r^{-1}  =  s_j \if \pi_r(\al_i)=\al_j,\  0\le j\le n.
&\eqnu
%&(1.6)
}
$$

We extend the notion
of the length to $W^b$.
Given $\nu\in\nu_R,\  r\in O^*,\  \tw \in W^a$, and a reduced
decomposition $\tw\ =\ s_{j_l}...s_{j_2} s_{j_1} $ with respect to
$\{s_j, 0\le j\le n\}$, we call $l\ =\ l(\hw)$ the {\it length} of
$\hw = \pi_r\tw \in W^b$. Setting
$$
\eqalign{
\la(\hw) = &\{ \tal^1=\al_{j_1},\
\tal^2=s_{j_1}(\al_{j_2}),\
\tal^3=s_{j_1}s_{j_2}(\al_{j_3}),\ldots \cr
&\ldots,\tal^l=\tw^{-1}s_{j_l}(\al_{j_l}) \},
}
\eqnu
$$
\label\tal\eqnum*
we introduce the {\it partial lengths}:
$$
l= \sum_\nu l_\nu,\  l_\nu = l_\nu(\hw)=|\la_\nu(\hw)|,
$$
where $|\ |$  denotes the  number of elements,
$$
\eqalignno{
&\la_\nu(\hw) = \{\tal^{m},\ \nu(\tal^{m})= \nu(\tal_{j_m})= \nu\},
1\le m\le l,
&\eqnu
%&(1.7)
}
$$
for $\nu([\al,k]) \equal \nu_{\al}$.

To check that these sets do not depend on the choice of the
reduced decomposition let us use the affine
action of $W^b$ on $z \in \R^n$:
$$
\eqalign{
& (wb')\langle z \rangle \ =\ w(b+z),\ w\in W, b\in B,\cr
& s_{\tal}\langle z\rangle\ =\ z - ((z,\al)+k)\al^\vee,
\ \tal=[\al,k]\in R^a.
}
\eqnu
% \eqno(1.8)
$$
and the affine Weyl chamber:
$$
\eqalignno{
&C^a\ =\ \bigcap_{j=0}^n L_{\al_j},\ L_{\tal}=\{z\in \R^n,\
(z,\al)+k>0 \}.
}
$$
\proclaim{Proposition}% 1.2 }
$$
\eqalign{
\la_\nu(\hw)\ & =\ \{\tal\in R^a, \ \hw^{-1}\langle  C^a \rangle
\not\subset L_{\tal}, \ \nu({\tal})=\nu \} \cr
& =\ \{\tal\in R^a, \ l_\nu( \hw s_{\tal}) < l_\nu(\hw) \}.}
\eqnu
% \eqno(1.9)
$$
\endproclaim
\proofbox

We mention that
$$
\eqalign{
&l(\ \hw s_{\tal\{1\}}...s_{\tal\{p\}}\  )\  >\
l(\ \hw s_{\tal\{1\}}...s_{\tal\{p+1\}}\  ), \if \cr
& \tal\{q\}\equal \tal^{m_q}, \ l\ge m_1> m_2>...>m_p>m_{p+1}\ge 1 .}
\eqnu
% \eqno(1.10)
$$
\label\decom\eqnum*

\proclaim{Proposition }%1.3}
 Each of the following conditions
for $x,y \in W^b$
is equivalent to the relation $ l_\nu(xy)=l_\nu(x)+l_\nu(y)$:
$$
\eqalign{
& a)\ \la_\nu(xy) = \la_\nu(y) \cup
y^{-1}(\la_\nu(x)) ,\ b)\  y^{-1}(\la_\nu(x)) \subset R^a_+ \cr
& c)\  \la_\nu(y) \subset \la_\nu(xy),
 \quad\  d)   y^{-1}(\la_\nu(x)) \subset \la_\nu(xy).
}
\eqnu
%\eqno(1.11)
$$
\endproclaim
\proofbox

\proclaim{Proposition }%1.4}
{i)  }
In the above notations,
$$
\eqalign{
&  \la(b') = \{ \tal, \al \in R_+,
( b, \al ) > k\ge 0\} \cup
\{ \tal, \al \in  R_-, ( b, \al ) \ge k>0 \},
 \cr
&  \la(\pi_b^{-1}) = \{ \tal,
-( b, \al )>k\ge 0 \}, \where \tal=[\al,k]\in R^a_+, b\in B.
}
\eqnu
%&(1.12)-(1.13)
$$
\label\la\eqnum*

{ii)} If $\hw \in b'W$ (i.e. $ \hw\langle 0\rangle = b$) then
$\hw\ =\ \pi_b w \for w\in W$ such that $l(\hw)=l(\pi_b)+l(w)$ .
Given $b \in B$, this  property (valid for any $\hw $ taking $0$
to $b$)  determines $\pi_b$ uniquely.
\endproclaim \proofbox
\label\piom\theoremnum*

 Relation (\ref\la)  gives the
following useful formulas:
$$
\eqalign{
&l_\nu(b')\ =\ \sum_{\al} |(b,\al)|,  \al\in R_+, \nu_\al=\nu \in \nu_R, \cr
&l_\nu(b'_+)\ =\ 2(b,\rho_\nu), \when b \in B_+.}
\eqnu
%\eqno(1.14)
$$
Here $|\ | = $
absolute value. Moreover
$$
\eqalignno{
& \la(\om_{b'_+}) = \{ \al \in R_+, ( b_+, \al ) > 0 \}
\for b_+ \in B_+.
&\eqnu
%&(1.15)
}
$$

 We set
$$
\eqalignno{
&c \preceq b, b\succeq c \for b, c\in B\iif b-c \in A_+.
&\eqnu
%&(1.16)
}
$$
%and use $\prec, \succ$ respectively if $b \neq c$.
Given $b\in B$, let $b_+= w_+^{-1}(b) \in B_+$ for  $w_+$
from  Definition 1.1.
The set
$$
\eqalign{
&\si^{\vee}(b)\equal \{c\in B, w(c)\preceq
b_+  \for \hbox{any\ } w\in W\}
}
%\cr
%&\si^{\vee}_0(b)\equal \{c\in B, w(c)\prec
%b_+  \for \hbox{any\ } w\in W\}}
\eqnu
%\eqno(1.17)
$$
is $W$-invariant (which is evident) and convex.
By {\it convex}, we mean that if
$ c,c^*= c+r\al^{\vee}\in \si^{\vee}(b)$
% (\in \si^{\vee}_0(b))$
for $\al\in R, r\in \Z_+$, then
$$
\eqalignno{
&\{c,\ c+\al^{\vee},...,c+(r-1)\al^{\vee},\ c^*\}\subset \si^{\vee}(b).
%( \subset \si^{\vee}_0(b)) .
&\eqnu
%&(1.18)
}
$$

We  also note that  $\si^{\vee}(b)$ contains the orbit
$W(b)$.

\proclaim{Proposition }%1.5 }
Given $ \hw \in W^b, \tal\in \la(\hw^{-1})$, let $b=\hw \langle 0
\rangle, \hw_*= s_{\tal}\hw , b_* = \hw_* \langle 0\rangle $. Then
$b_* \in \si^{\vee}(b)$,
$$
\eqalign{
&\ell( \hw_*) < \ell(\hw)
\if b_*\neq b, \where \ell( \hw)=\ell(b')
\equal l(\pi_b),\cr
&\ell(b'_+) < \ell(\hw) < l(b')=\ell(b'_-) \if
b_+\neq b\neq b_- \equal -b_+.}
\eqnu
%\eqno (1.19)
$$
Moreover $\ell(\hw_*)< \ell(\hw)$
for $\hw_*=  s_{\tal\{p\}}...s_{\tal\{1\}}\hw$, where we
take any sequence (\ref\decom) for $\hw^{-1}$ (instead of $\hw$)
such that $\ell(s_{\tal\{1\}}\hw) < \ell(\hw)$.
% The  elements $b_*$ for such $\hw_*$
%corresponding to
% $b=b_+$ constitute  the set $\si^{\vee}_0(b)$.
\endproclaim
\proofbox
\label\ell\theoremnum*

%
%
%		Section 2
%
%
%
%
%\vskip 10pt
\section{ Affine R-matrices}
We fix an arbitrary $\C$-algebra $\FF$. We introduce
$\FF$-valued (abstract) affine R-matrices following [C4] (see also
[C3]), define the   monodromy cocycle of the corresponding
Knizhnik-Zamolodchikov equation, and  extend the latter
to  a new affine R-matrix. Starting with the basic
trigonometric ones from
[C4], this construction gives  R-matrices of elliptic type.
We will use the
notations from Section 1. Let us denote $\R\tal + \R\tbe \subset \R^n $
by $\R \langle \tal, \tbe \rangle \for \tal,\tbe \in R^a $.

\proclaim{Definition} %2.1}
{ a) } A set $G = \{ G_{\tal} \in \FF, \tal \in
R^a_+ \} $ is an R-matrix if
$$
\eqalignno{
& G_{\tal} G_{\tbe} = G_{\tbe} G_{\tal},
 &\eqnu %&(3.1)
}
$$
\label\two\eqnum*
$$
\eqalignno{
& \qquad \ \ G_{\tal} G_{\tal+\tbe} G_{\tbe} =
G_{\tbe} G_{\tal+\tbe} G_{\tal},
&\eqnu
\cr
\label\three\eqnum*
& G_{\tal} G_{\tal+\tbe} G_{\tal+2\tbe} G_{\tbe} = G_{\tbe} G_{\tal+2\tbe}
G_{\tal+\tbe} G_{\tal},
&\eqnu %&(3.3)
\label\four\eqnum*
}
$$
$$
\eqalignno{
& G_{\tal} G_{3\tal+\tbe} G_{2\tal+\tbe} G_{3\tal+2\tbe}
G_{\tal+\tbe} G_{\tbe} =
G_{\tbe} G_{\tal+\tbe} G_{3\tal+2\tbe} G_{2\tal+\tbe}
G_{3\tal+\tbe} G_{\tal} &\eqnu %&(3.4)
}
$$
\label\six\eqnum*
under the assumption that $\tal, \tbe \in R^a_+ $ and
$$
\eqalignno{
&\R \langle \tal, \tbe \rangle \cap R^a = \{ \pm \ga \}, \ga \ \hbox{
 runs over all the indices}  &\eqnu %&(3.5)
}
$$
\label\belong\eqnum*
in the corresponding identity.

{ b) } A non-affine R-matrix $ \{ G_{\al} \in \FF, \al \in
R_+ \} $ has to obey the same relations for $\al, \be \in R_+$.
 A closed R-matrix (or a closure of the above $G$) is a
set $ \{ G_{\tal} \in \FF, \tal \in R^a \} $   (extending $G$ and)
satisfying relations (\ref\two - \ref\six) for arbitrary (maybe negative)
$ \tal, \tbe \in R^a $ such that the corresponding condition
(\ref\belong) is
fulfilled. Non-affine closed R-matrices are defined for $\al\in R$.
\endproclaim
\proofbox
\label\G\theoremnum*

The condition (\ref\belong) for identity (\ref\two) means that
$$
\eqalignno{
& (\tal, \tbe) = 0 \and \R \langle \tal, \tbe \rangle \cap R^a
 = \{ \pm \tal,
\pm \tbe \},   &\eqnu %&(3.6)
}
$$
i.e. there exists  $\tw \in W^a$ such
that $ \tal = \tw(\al_i), \tbe = \tw(\al_j)$
for simple $ \al_i \ne \al_j (0\le i,j \le n) $ disconnected in $\Ga^a$.
The corresponding assumptions for (\ref\three-\ref\six)
 give that $\tal, \tbe $ are
simple roots of a certain two-dimensional root subsystem in $R^a$ (or
$R$) of type $A_2, B_2, G_2$. Here $\tal, \tbe$ stay for $\al_1 ,
\al_2$ in the notations from the figure of the systems of rank 2 from [B] .
One can represent them as follows:
$ \tal = \tw(\al_i), \tbe = \tw(\al_j)$ for
a proper $\tw$ from $W^a$ and joined  (neighbouring )
$\al_i, \al_j$.

We will use the following formal notations:
$$
\eqalignno{
&{}^{\hw}(G_{\tal}) =  G_{\hw(\tal)}, \quad {}^{\hw}(G_{\tal}G_{\tbe}) =
 G_{\hw(\tal)} G_{\hw(\tbe)}, ... ,  &\eqnu %&(3.10)
}
$$
where $\hw$ is from $ W^b$, the roots  $\tal, \tbe $ are from $R^a$.
 We do not assume here that  $W^b$
 acts on $\FF$. However one can consider $\{G_{\tal}\}$ as formal
symbols satisfying the relations of the Definition \ref\G and
$\FF$ as the free algebra generated by them over $\C$. Then indeed
the above formulas define an action.

\proclaim{Proposition}% 2.3}
 If $G$ is an affine R-matrix then there exists a
unique set $\{ G_{\hw}, \hw \in W^b \}$ satisfying the
(homogeneous 1-cocycle) relations
$$
\eqalignno{
&G_{xy} = {}^{y^{-1}}G_x G_y, \quad G_{s_j} = G_j \equal G_{\al_j} ,
G_{id} = 1, &\eqnu %&(3.12)
}
$$
where $0 \le j \le n ,\  x,y \in W^b \and l(xy) = l(x) + l(y) $.
\endproclaim
\proofbox

Given  a reduced decomposition
 $\hw= \pi_r s_{j_l}\cdots s_{j_1}$, $\ l=l(\hw), r\in O$,
 one has (see (\ref\tal)):
$$
\eqalignno{
&G_{\hw}=\hw G_{\tal^l}\cdots G_{\tal^1},\  \tal^1=\al_{j_1},
\tal^2=s_{j_1}(\al_{j_2}),
\tal^3=s_{j_1}s_{j_2}(\al_{j_3}),\ldots.%&(3.13)
&\eqnu
}
$$

 We always have the following
closures ({\it the unitary one} defined for invertible $G$
 and {\it the extension by 0}):
$$
\eqalignno{
&G_{-\tal }\ =\ G_{\tal}^{-1},\quad
G_{-\tal }\ =\  0, \quad \tal \in R^a_+ .%&(2.7)
&\eqnu
}
$$
If there exists an action of $W^a\ni \tw$ on $\FF$ such that
$$\tw(G_{\tal})\ =\ {}^{\tw}G_{\tal}\ =\ G_{\tw(\tal)}
\for \tal,\ \tw(\tal)\in R^a_+,$$
then the extension of $G$ satisfying  these relations for
all $\tw$ is well-defined and closed ({\it
the invariant closure}).

\proclaim{Proposition}% 2.2}
 Let us suppose that the non-affine R-matrix $G$ is closed and
$$
\eqalignno{
&G_{\al} G_{\be} = G_{\be} G_{\al} \quad \hbox{for \  long \  roots \
such \ that \ }
 (\al,\be)= 0. &\eqnu %&(3.7)
}
$$
\label\orth\eqnum*
In the case of $G_2$ we take all  orthogonal  roots in
(\ref\orth) and
 add  condition (\ref\three)  for
long $\al ,\be$. We call such an R-matrix extensible.
 Assuming
that the group $B \ni a $  operates on the algebra
$\FF \ni f $ (written $ f \to b(f) $) and
$$
\eqalignno{
&b( G_{\al} ) = G_{\al} \quad\hbox{whenever}\quad (b,\al) = 0,\
b \in B,
\al \in R,  &\eqnu %&(3.9)
}
$$
we set
$$
\eqalignno{
&{}^{b'}G_{\al } =  G_{\tal} \equal b( G_{\al} ),\if \tal = b'(\al) =
[\al,-(b,\al)] &\eqnu %&(3.15)
}
$$
for arbitrary $\al \in R, b \in B$.
Then $G_{\tal}$ are well-defined (depend on the
corresponding scalar products $(b,\al)$ only) and form a closed affine
R-matrix.

\endproclaim
\proofbox
\label\exten\theoremnum*

{}From now on we assume that affine $G$ is obtained by this
construction. We see that $G_{\hw}$  is the product of $G_{\tal}$
when $\tal$ runs over $\la(\hw)$. It does not depend
on the choice of the closure of $G$ if $\la_w$ contains no elements $\tal =
[\al,k] $ with $\al < 0$ (we call such $w \in W^b$ {\it dominant}).
Moreover,
 the elements $G_w$ for non-dominant $w$ vanish
if the  closure is the extension by 0.
If the closure  is unitary, then formulas (2.12) are
valid
for any $x,y \in W^b$.

\proclaim{Proposition}% 2.4}
 The elements from \ $B'_+ = \oplus _{i=1}^n \Z b'_i $ \
are dominant. Given \ $b,c \in B_+, \ l((b+c)^\prime ) = l(b') +
l(c'),  $ and
$$
\eqalignno{
&G_{b'+c'}\ = \ (-c)(G_{b'})\ G_{c'} = \ (-b)(G_{c'})\ G_{b'},
 &\eqnu %&(3.18)
\cr
&G_{b'_r}\ = \ G_{\om_r}, \where r\in O^*,\ G_{\th'}\ =\
G_{[\th,1]}G_{s_{\th}}.
&\eqnu %&(3.19)
}
$$
\endproclaim
\proofbox
\label\commut\theoremnum*

\proclaim{Definition}% 3.1}
The quantum Knizhnik- Zamolodchikov equation
 is one of the following equivalent systems of relations for an element
\ $\Phi \in \FF$ \ :
$$
\eqalign{
&i) \ b^{-1}_i\ (\Phi) \ = \ G_{b'_i}\ \Phi \ ,\ 1 \le i \le n \ ;
\cr
&ii) \ b^{-1}\ (\Phi) \ = \ G_{b'}\ \Phi \for \ \hbox{any} \ b \in B_+.
}
\eqnu
$$
\label\qkzeq\eqnum*
\endproclaim
\proofbox
\label\qkz\theoremnum*

Let us assume that
 there is an action of $W^b$ on $\FF$ making $G$ invariant
or/and $\Phi$ is a series (or any expression)
 in terms of $\{G_{\tal}\}$.
For invertible $\Phi$, the
  (homogeneous) {\it monodromy cocycle} $\{\g_w, w\in W\}$
is determined by  the relations:
$$
\eqalign{
&\g_i=\g_{s_i} \equal {}^{w^{-1}}(\Phi^{-1})G_i\Phi \for 1\le i\le n,
\ \g_{id} = 1,
\cr
&\g_{xy} = {}^{y^{-1}}\g_x \g_y \hbox{\ \ whenever\ \ }
l(xy) = l(x) + l(y).
}
\eqnu %\eqno(3.12)
$$
\label\monod\eqnum*
The existence results from the relations for $\{\g_i\}$
from Definition \ref\G.

Let us suppose that the limits
$$
\Phi \ =\ \lim_{b\to +\infty} b(G_{b'}),\
{}^w\Phi \ =\ \lim_{b\to +\infty}  w(b)({}^w G_{b'}),\ w\in W,
\eqnu
%\eqno(3.12)
$$
\label\phi\eqnum*
are well-defined
 in the algebra $\FF$ together with  their finite products
  as all $\{k_i\}$
from the decomposition $b=\sum_{i=1}^n k_ib_i \in B_+$
tend to $\infty$. Then
$\Phi$ is a solution of (\ref\qkzeq). We assume that the action of $B$
is continuous in the topology of $\FF$.

\proclaim {Proposition}
The following products
$$
\g_{\al} \equal\cdots G_{[-\al,-2]}^{-1} G_{[-\al,-1]}^{-1}
G_{\al} G_{[\al,-1]} G_{[\al,-2]}\cdots
\eqnu
%\eqno(3.12)
$$
are convergent and form a  non-affine closed extensible
 R-matrix ($\al\in R$).
When $\al=\al_i$ they coincide with $\g_i$ from (\ref\monod)
for the above $\Phi$.
\endproclaim
\proofbox
\label\convergent\theoremnum*

\proclaim{Theorem}
In the above setup, assume
that there is another continuous action of the group $B$ (written
$\bio(\ )$) on the algebra
$\FF $ such that
$$
\eqalign{
\bio( G_{\al} ) = G_{\al} &\quad\hbox{whenever}\quad (b,\al) = 0,\
b \in B,
\al \in R, \cr
\bio a'\  = \ a'\bio &\quad\hbox{whenever}\quad (b,a) = 0,\
a,b \in B.
}
\eqnu %\eqno(3.9)
$$
Then
$$
\eqalignno{
&\gio_{\tal} \equal \bio( \g_{\al} ),\if \tal = b'(\al) =
[\al,-(b,\al)] &\eqnu %&(3.15)
}
$$
for  $\al \in R, b \in B$
 are well-defined (depend on  $(b,\al)$ only) and form a closed affine
R-matrix.
\endproclaim
\proofbox
\label\extended\theoremnum*

Examples will be considered in Section 4.

%
%		Section 3
%
%
%
%
%\vskip 10pt
\section{  Double affine Hecke algebras}

We denote the  least common order of the elements of $\Pi$ by $m \
(m=2 \for D_{2k}$, otherwise $m=|\Pi|)$. Later on $ \C_{\de,q}$
 means the field of rational
functions in parameters $\{\de^{1/m}, q_\nu^{1/2} \}, \nu \in \nu_R$.
We will not distingwish $b$ and $b'$ in this and the next section.
Let
$$
\eqalignno{
&   q_{\tal} = q_{\nu(\tal)},\ q_j = q_{\al_j},
\where \tal \in R^a, 0\le j\le n.
%&(2.1)}
&\eqnu}
$$
%\label\(2.1)\eqnum*
Setting
$$x_i=\exp({b_i}),\  x_b=\exp(b)=
\prod_{i=1}^n  x_i^{k_i}
\for b=\sum_{i=1}^n k_i b_i,$$
$\C_\de[x] = \C_\de[x_b]$ means the algebra of
polynomials in terms of $x_i^{\pm 1}$
with the coefficients depending
on $\de^{1/m}$ rationally.
We will also use
$$
\eqalignno{
&X_{\tb} = \prod_{i=1}^nX_i^{k_i}\de^{ k} \if \tb=[b,k],\
b=\sum_{i=1}^nk_i b_i\in B,\ k \in {1\over m}\Z,
%&(2.2)}
&\eqnu}
$$
\label\Xde\eqnum*
where $\{X_i\}$ are independent variables which act in  $\C_\de[x]$
naturally:
$$
\eqalignno{
& X_{\tb}(p(x))\ =\ x_{\tb} p(x),  \where x_{\tb}\equal x_b\de^{ k},\
  p(x) \in
\C_\de [x].
%&(2.3)}
&\eqnu}
$$
\label\X\eqnum*
The elements $\hw \in W^b$ act in $\C_{\de}[x],\  \C_{\de}[X]=
\C_\de[X_b]$ by the
formulas:
$$
\eqalignno{
&\hw(x_{\tb})\ =\  x_{\hw(\tb)}, \ \ \hw X_{\tb}\hw^{-1}\ = \
X_{\hw(\tb)}.
%&(2.4)}
&\eqnu}
$$
 In particular:
$$
\eqalignno{
&\pi_r(x_{b})\ =\ x_{\om^{-1}_r(b)}\de^{(b_{r^*},b)}
\for \al_{r^*} \equal \pi_r^{-1}(\al_0), \ r\in O^*.
%&(2.5)}
&\eqnu}
$$
\label\pi\eqnum*
We set $([a,k],[b,l])=(a,b)$ for $a,b\in B,\
[\al,k]^\vee= [\al^\vee,k]$, and $a_0=\al_0$.

\proclaim{Definition }%2.1}
 The  double  affine Hecke algebra $\HH\ $
(see [C1,C2])
is generated over the field $ \C_{\de,q}$ by
the elements $\{ T_j,\ 0\le j\le n\}$,
pairwise commutative $\{X_b, \ b\in B\}$ satisfying (\ref\Xde),
 and the group $\Pi$ where the following relations are imposed:

(o)\ \  $ (T_j+q_j^{1/2})(T_j-q_j^{-1/2})\ =\ 0,\ 0\ \le\ j\ \le\ n$;

(i)\ \ \ $ T_iT_jT_i...\ =\ T_jT_iT_j...,\ m_{ij}$ factors on each side;

(ii)\ \   $ \pi_rT_i\pi_r^{-1}\ =\ T_j \if \pi_r(\al_i)=\al_j$;

(iii)\  $T_i^{-1}X_b T_i^{-1}\ =\ X_b X_{a_i}^{-1} \if (b,\al_i)=1,\
1 \le i\le  n$;

(iv)\  $T_0^{-1}X_b T_0^{-1}\ =\ X_{s_0(b)}\ =\ X_b X_{\th}\de^{-1}
\if (b,\th)=-1$;

(v)\ \ $T_iX_b\ =\ X_b T_i$ if $(b,\al_i)=0 \for 0 \le i\le  n$;

(vi)\ $\pi_rX_b \pi_r^{-1}\ =\ X_{\pi_r(b)}\ =\ X_{\om^{-1}_r(b)}
\de^{(b_{r^*},b)},\ r\in O^*$.
\endproclaim
\proofbox
\label\double\theoremnum*

Given $\tw \in W^a, r\in O,\ $ the product
$$
\eqalignno{
&T_{\pi_r\tw}\equal \pi_r\prod_{k=1}^l T_{i_k},\where
\tw=\prod_{k=1}^l s_{i_k},
l=l(\tw),
%&(2.6)}
&\eqnu}
$$
\label\Tw\eqnum*
does not depend on the choice of the reduced decomposition
(because $\{T\}$ satisfy the same ``braid'' relations as $\{s\}$ do).
Moreover,
$$
\eqalignno{
&T_{\hv}T_{\hw}\ =\ T_{\hv\hw}\  \hbox{ whenever}\
 l(\hv\hw)=l(\hv)+l(\hw) \for
\hv,\hw \in W^b.
 %&(2.7)}
&\eqnu}
$$
\label\TT\eqnum*
  In particular, we arrive at the pairwise
commutative elements (use (\ref\TT) and Proposition \ref\commut):
$$
\eqalignno{
& Y_{b}\ =\  \prod_{i=1}^nY_i^{k_i} \if
b=\sum_{i=1}^nk_ib_i\in B,\where
 Y_i\equal T_{b_i}.
%&(2.8)}
&\eqnu}
$$

\proclaim{Proposition}% 2.2}
$$
\eqalign{
&T^{-1}_iY_b T^{-1}_i\ =\ Y_b Y_{a_i}^{-1} \if (b,\al_i)=1,
\cr
& T_iY_b\ =\ Y_b T_i \if (b,\al_i)=0, \ 1 \le i\le  n.}
%\eqno(2.9)
\eqnu
$$
\endproclaim
\proofbox

We note that the above relations are equivalent to the following:
$$
\eqalignno{
& T_i^{-1}Y_iT_i^{-1} =  Y_iY_{a_i}^{-1},
\ T_iY_j=Y_jT_i\for 1\le i\neq j\le n .
 %&(2.10)}
&\eqnu}
$$
\label\TY\eqnum*

\proclaim {Proposition}
The following three maps define automorphisms of \HH:
$$
\eqalign{
 \vep: &\ X_i \to Y_i^{-1},\ \  Y_i \to X_i^{-1},\  \ T_i \to T_i^{-1}, \cr
&q_\nu \to q_\nu^{-1},\
\de\to \de^{-1},\ 1\le i\le n,\ \nu\in \nu_R,
}
\eqnu
$$
\label\om\eqnum*
$$
\eqalign{
 \tau: \ &X_i \to X_i,\ \ Y_i \to X_iY_i\de^{-c_i},\
\ T_i \to T_i, \cr
&q_\nu \to q_\nu,\
\de\to \de,\ c_i=(b_i,\rho_2)/(1+(\th,\rho_2)),\cr
 \om: \ &X_i \to Y_i,\ \ Y_i \to Y_i^{-1}X_iY_i\de^{2c_i},\
\ T_i \to T_i, \cr
&q_\nu \to q_\nu,\
\de\to \de,\ 1\le i\le n,\ \nu\in \nu_R.
}
\eqnu
$$
\endproclaim
\proofbox
\label\dual\theoremnum*

This theorem can be either deduced from the topological interpretation
of the double braid group from [C1] (defined by relations (i)-(vi),
Definition \ref\double) or checked by direct consideration.
One has to add proper roots of $\de$ to \HH\ when defining
$\tau, \om$ (corresponding to the standard generators of
$SL_2(\Z)$ in the case $q=1=\de$).
The first automorphism can be interpreted
as the self-duality of the (formal) zonal Fourier transform
in the difference case (see [DJO] for the differential rational
case). Conjecturally the symmetric polynomials of
the  images of $\{Y_1,\ldots,Y_n\}$ with respect
to the action of the proper products of $\tau,\om$
generate certain difference
analogs of the so-called $W_\infty$-algebras from conformal
field theory.

The {\it Demazure-Lusztig operators} (see
[KL, KK, C1], and [C5] for more detail )
$$
\eqalignno{
&\hT_j\  = \  q_j ^{-1/2} s_j\ +\
(q_j^{-1/2}-q_j^{1/2})(X_{a_j}^{-1}-1)^{-1}(s_j-1),
\ 0\le j\le n.
%&(2.11)}
&\eqnu}
$$
act   in $\C_{\de,q}[x]$ naturally.
We note that only $\hT_0$ depends on $\de$:
$$
\eqalign{
&\hT_0\  =  q_0^{-1/2}s_0\ +\ (q_0^{-1/2}-q_0^{1/2})
(\de^{-1} X_{\th} -1)^{-1}(s_0-1),\cr
&\where
s_0(X_i)\ =\ X_iX_{\th}^{-(b_i,\th)}\de^{(b_i,\th)}.
}
%\eqno(2.12)
\eqnu
$$

\proclaim{Theorem }%2.3}
{ i)  } The map $ T_j\to \hT_j,\ X_b \to X_b$ (see (\ref\Xde,\ref\X)),
$\pi_r\to \pi_r$  (see (\ref\pi)) induces a $ \C_{\de,q}$-linear
homomorphism from \HH\ to the algebra of linear endomorphisms
of $\C_{\de,q}[x]$.

{ ii) } This representation is faithful and
remains faithful when   $ \de,q$ take  any non-zero
values assuming that
 $\de$ is not a root of unity.
 Elements $H \in \HH\ $ and their images $\hH$ have
the unique decompositions
$$
\eqalignno{
&H =\sum_{b\in B, w\in W } Y_b h_{b,w}  T_w,\ \
h_{b,w} \in \C_{\de,q}[X].
%&(2.13)
&\eqnu\cr
&\hH\ = \sum_{b\in B, w\in W} b g_{b,w}  w\ =\
\sum_{b\in B, w\in W}  b(g_{b,w}) \ b  w,
%&(2.14)
&\eqnu
}
$$
where  $g_{b,w}$ belong to  the field $\C_{\de,q}(X)$
of rational  functions in
$\{X_1,...,X_n\}$.
\endproclaim
\proofbox
\label\faith\theoremnum*

\proclaim{Proposition} %2.4  }
i) Given $b\in B$ and $\hw = \pi_b\om,\  \om \in W$,
$$
\eqalignno{
&\hT_{\hw} = <\hT_{\hw}>
 + \sum_{b_*, w\in W}  b_* g_{b_*,w} w,
%&(2.15)}
&\eqnu}
$$
summed over $ b_*\in \si^\vee(b)$ such that
$  \ell(b_*) < \ell(b)$,
where $g_{b,w} \in\C_{\de,q}(X)$,
$$
\eqalign{
&<\hT_{\hw}>\  = \
 \prod _{\tal \in \la(\pi_b^{-1})}
{q_{\tal}^{1/2} X_{\tal^\vee}- q_{\tal}^{-1/2}\over
X_{\tal^\vee}- 1}
b\om_b^{-1}  \hT_{\om}.
}
%\eqno (2.16)
\eqnu
$$

ii) If $ b \in B_+$,
%then $T_{b}=Y_b$ and all $\ b_*$ belong to $ \si_0^\vee(b)$.
%Moreover
then $\pi_{-b}=-b$, $Y_{-b}= T_{b}^{-1}$, and
$$
\eqalign{
&\hY_{-b} =  <\hY_{-b}> + \sum_{b_*, w\in W} b_* g_{b_*,w}  w,
\quad\  b\neq b_* \in \si^\vee(b),
  \cr
&<\hY_{-b}> \  = \
 \prod _{\tal \in \la(b) }
{q_{\tal}^{1/2} X_{\tal^\vee}- q_{\tal}^{-1/2}\over
X_{\tal^\vee}- 1} (-b).
}
%\eqno(3.5)
\eqnu
$$
%Here we omit the condition
%$ \ell(b)  > \ell(b_*)$  because it is valid for any
%$b\neq b_*\in \si^\vee(b)$ (see Proposition \ref\ell).
\endproclaim
\proofbox
\label\leading\theoremnum*
%
%		Section 4
%
%
%
%
%\vskip 10pt
\section{ Difference-elliptic operators }
% Identifying
%\HH\  with its image with respect to $\hze$,
%we write $T,Y$ instead of $\hT,\hY$.

%Let $\h_X, \ \h_Y$ be  the {\it affine Hecke algebras} generated
%over $\C_q$ by
%abstract $\{T_i, 1 \le i \le n\}$ and pairwise commutative
%$\{X_i\}$, $\{Y_i\}$
%satisfying relations (o,i,iii,v) from Definition \ref\double
% (for $1\le i,j \le n$)
%and (\ref\TY). We may
%regard them as subalgebras of \HH\ ( which is
%possible thanks to Theorem \ref\faith, ii)).
%It also can be done when

The above considerations lead to the following affine
R-matrix:
$$
\eqalignno{
&G_{\tal;q}  = G_{\tal} = 1+(q_{\tal}-1)
(X_{\tal^\vee} -1)^{-1}(s_{\tal}-1) =
G_{\tal;q^{-1}}^{-1},\ \tal\in R^a.
&\eqnu %&(2.17)
}
$$
Given  a reduced decomposition
 $\hw= \pi_r s_{j_l}\cdots s_{j_1}$, $\ l=l(\hw), r\in O$,
 one has (see (\ref\Tw)):
$$
\eqalign{
&\hT_{\hw}\ =\ \hw \prod_\nu q_\nu^{-l_\nu(\hw)/2}
G_{\tal^l}\cdots G_{\tal^1},\cr
  &\tal^1=\al_{j_1},
\tal^2=s_{j_1}(\al_{j_2}),
\tal^3=s_{j_1}s_{j_2}(\al_{j_3}),\ldots.
}
\eqnu%\eqno(2.18)
$$
We will apply the procedure of  Proposition \ref\convergent
to the  $G$. From now on  we fix $q\neq 0$.

\proclaim {Definition}
 Given  ${\hbox{\it \ae}}>0,\ M>1$, let
$$
\eqalign{
\Xi_{\hbox{\it \ae}}(M)\ =\ \{&\ (x,\de)
\hbox{\ such\ that}\ \
 |\de| \le \exp({-\hbox{\it \ae}}),\cr
 & | (x_{\al}\de^k-1)^{-1}| <  M >
|x_{\al}|\ \},
}
\eqnu
$$
 for all $k\in \Z,\ \al\in R$, where
$x=(x_1,\ldots,x_n)\in \C^n$.
We denote by \FF\ the algebra of the
 series
$$
f=\sum_{\hw}f_{\hw}(x,\de)\hw \for
 \hbox{\ scalar\ }f_{\hw},\ \hw\in W^b,
\eqnu
$$
\label\f\eqnum*
 satisfying the following
condition.
There exists a constant $c_f>0$ (depending on $f$) such that
for any $M>1,\hbox{\it \ae}$, and $\hbox{\it \ae}\ge\ep>0$ the
functions $\{f_{\hw}\}$ are continuous in $\Xi_{\hbox{\it \ae}}(M)$
and the series
 (\ref\f) is convergent with
respect to the norm
$$
\eqalign{
&||f|| \equal
\sum_{\hw}\sup\{ | f_{\hw} | \hbox{\ in\ }
\Xi_{\hbox{\it \ae}}(M)\} ||\hw||,\cr
&\where
||\hw||\  = \ \exp (c_f(\hbox{\it \ae}-\ep)l(\hw)).
%\eqno 2.15
}\eqnu
$$
Here $l(\hw)$ is the length of $\hw\in W^b$,
$|\ |$  the absolute value.
\endproclaim
\proofbox
\label\domain\theoremnum*

One can follow [C3],Proposition 2.5 to check that \FF\ is
really an algebra.
%The elements from this algebra
%considered in  any given finite dimensional representation
%of $W^b$
%are
%absolutely convergent series in $\Xi_{\hbox{\it \ae}}(M)$
%for sufficiently big ${\hbox{\it \ae}}$ and arbitrary $M$.

 Let us add two more variables $\xi,\ze$ and one more
parameter $t$ to $\C_{\de} [x]$  and extend
the action of $b\in B$ and $W^b$
setting
$$
\eqalignno{
& b(\xi) = \xi x_{b} \de^{-(b,b)/2},\
 w(\xi) = \xi, \ w\in W, \ b(\ze) = \ze.
%&(2.3)}
&\eqnu
}
$$
\label\b\eqnum*
Replacing $\de$ by $t$ in formulas (\ref\Xde-\ref\pi)
 and $\xi$ by $\ze$,
we introduce another
action of $b\in B$:
$$
\eqalignno{
&\bio (\ze) = \ze x_{b} t^{-(b,b)/2},\
 w(\ze) = \ze, \ w\in W, \ \bio(\xi) = \xi.
%&(2.3)}
&\eqnu
}
$$
\label\bi\eqnum*

The group $\w$ of automorphisms of $\C_{\de,t}[x,\xi,\ze]$
generated by $W^b$, $B^\iota$ is isomorphic
to the following central extension of the
the semi-direct product $ W\lsmash (B\oplus \Bio)$
by $\Z=\{d^l, l\in\Z\}$.
The action of $W$ on $B, \Bio$ by
conjugation is given by the same
formulas  as in Section 1. However
 $$
\eqalign{
& \aio b\ =\ b\aio  d^{(a,b)} \for a,b\in B, \where\cr
&d(x_{b}) = x_{b},\ d(\xi)\ \equal \xi t^{-1},\
d(\ze)\ \equal \ze \de.
}
%\eqno(2.5)}
\eqnu
$$
\label\central\eqnum*
Formulas (\ref\b,\ref\bi)  will
not be applied in this paper but the last one will.
We note that the action of $\w$ in
$C[x]$ remains  faithful when
 $\de,t$ are considered as  numbers assuming that
$t^r\de^s=1$ iff $r=0=s$ for integral $r,s$.
We will call $\w$ {\it 2-extended Weyl group} due to K. Saito.

%We will use the notations $l,\ l_\nu$
%for the lengths of the projections of the elements of $\w$
%onto $ W^b$ and $l^\iota,\ l^\iota_\nu$ for the
%lengths of the projections
%onto $W^b_\iota \equal W\lsmash \Bio$.

Thanks to (\ref\central) we can construct
the element $\Phi$ from (\ref\phi) and the
R-matrix $\g$  (Proposition \ref\convergent)
which belong to \FF\. Let us extend it to the
 affine closed  $ W^b_\iota$-invariant
R-matrix $\{\gio_{\tal}\}$, where the action is
by conjugation relative to
 (\ref\bi) (see Theorem
\ref\extended). This R-matrix belongs to the
algebra $\FF^{\#}$ of finite sums
 $\sum_{\hw,r}F_{\hw,r}(x,\de)\hw^\iota d^r$, where
 $F_{\hw}\for \hw\in W^b,\ r\in (1/m)\Z$ are finite products
of series (\ref\f) "shifted" by any $\bio$.

Introducing $\gio_{\hw}$ for
$\hw\in W^b$, we set
$$
\eqalign{
&\t_{\hw}\ =\ \prod_\nu q_\nu^{-l_\nu(\hw)/2}
\hw \gio_{\hw} \for \hw\in W^b,\cr
&\y_b \ =\ \t_{b},\ \y_{a-b}\ =\ \y_a\y_b^{-1},\for
a,b\in B_+.
}
\eqnu
$$

\proclaim{Theorem}
The map $\phi$ sending
$$
\eqalign{
&T_i \to \t_i\equal \Phi^{-1} T_i \Phi \for 1\le i\le n,\cr
&Y_i \to \y_i\equal \prod_\nu q_\nu^{-(\rho_\nu,b_i)}
\bio_i \g_i\cr
&X_i \to \x_i\equal \Phi^{-1} (Y_i) \Phi=
\prod_\nu q_\nu^{-(\rho_\nu,b_i)}
b_i,\cr
&\de \to d^{-1}\De,\ \De\equal
q_0^{-1}\prod_\nu q_\nu^{-(\rho_\nu,\th)}
}
\eqnu
$$
\label\critical\eqnum*
can be extended to a homomorphism of \HH\ .
\endproclaim
\proofbox
\label\main\theoremnum*

% The algebra of $W$-invariant elements  in the $\C[Y]$
%is denoted by
%$\C[Y]^W$. Here (and in similar cases) $x$ will be replaced
%by $X$ and other letters without more comment.
%We will  use that $\C[Y]^W$ is the center of $\h_Y$. The
%same of course holds for $\C[X]^W$ and $\h_X$. This property is
%due to Bernstein (see e.g. [L1], [C3]).

Let $p(x_1,\ldots,x_n)$ belong to the algebra $\C[x]^W$
of $W$-invariant elements  in the $\C[x]$.
We set
$$
\eqalign{
&\l_p=p(\y_1,....,\y_n) =\sum_{b\in B,\hw\in W^b,  r}
  f_{b,\hw,r} \bio \hw d^r,\ r\in {1\over m}\Z, \cr
&(\l_p)_{red} = \sum_{b,\hw,r}
 f_{b,\hw,r}\bio d^r,\  L_p\equal \sum_{b,\hw,r}
 f_{b,\hw,r}\bio \De^r.
}
\eqnu%\eqno(3.6)
$$
\label\red\eqnum*

\proclaim{Theorem}% 3.3 }
The operators $\{ L_p, \for p\in \C[y]^W\}$
are pairwise commutative,  $W^b$-invariant (i.e $\hw L_p \hw^{-1}=$
$L_p$ for all $\hw\in W^b$).
Given $p$, there exists
a finite set $\La\subset \Z$ such that
the coefficients $g_b= \sum_{\hw\in W^b, r} f_{b,\hw,r}\De^r$
 from the decomposition
 $L_p \ =\ \sum_{b\in B}
  g_{b}\bio$
 are absolutely convergent series in
$$
\eqalign{
&\Xi_{\hbox{\it \ae},t}(M;\La)\ =\ \{\ (x,\de,t),
\   t\neq 0,\ |\de| \le \exp({-\hbox{\it \ae}}),\cr
 & | (x_{\al}\de^k t^l-1)^{-1}| <  M >
|x_{\al}|,\where k\in \Z,\ l\in \La\  \}
}
\eqnu
$$
for  any  $M>1, \hbox{\it \ae}>0$.
They satisfy the following
periodicity condition:
$$
a(g_b(x))\ =\ \De^{(a,b)}g_b(x)\for a,b\in B.
$$
\endproclaim

{\it An outline of the proof.}
 First, the operators $\l_p$ are pairwise commutative.
Then $\C[X]^W$ belongs to the center of \HH\ when $\de=1$.
which follows from Definition \ref\double and Theorem \ref\faith.
The same holds true for
$\C[Y]^W$ because of Proposition \ref\dual (use $\vep$).
Hence (see Theorem \ref\main) the  above operators
belong to the center of the image of \HH\ relative to
$\phi$. It gives the $W^b$-invariance which directly results
in the statements of the theorem.
\proofbox

We mention without going into detail
 that  $\{L_p\}$ are formally self-adjoint
with respect to the "elliptic" Macdonald pairing and
preserve some remarkable finite dimensional subspaces
(cf. [C3]). Let us demonstrate that  the above construction
gives also  the commutativity of the operators from [R]
(generalized to
arbitrary root
systems). We set
$$
\eqalign{
&\l_p\ =\ \sum_{a,b\in B,w\in W,  r}
  f_{a,b,w,r} a\bio w d^r,\ r\in {1\over m}\Z, \cr
&\l_p^0 = \sum_{a,b,w,r}
 f_{a,b,w,r} a\bio d^r,\  L_p^0\equal \sum_{b,w}
 f_{0,b,w,0}\,\bio.
}
\eqnu%\eqno(3.6)
$$
\label\red0\eqnum*
Then $\l_p^0\l_g^0=\l_{pg}^0$
for any $W$-invariant $p,g$.

Considering  $p_r=\sum_{w\in W} w(b_r)$ for $r\in O^*$
and following Proposition 3.4 from
[C2],
we see that $\l_r^0\equal\l_{p_r}^0=\sum_{a,b,w}
 f_{a,b,w,0}\, a\bio,$ where
$(a,b)> 0$ for any non-zero pairs $a,b$.
Hence
$$
\eqalign{
&\l_s^0\l_r^0= \l_{p_s p_r}^0= \l_s^0(\sum_{a,b,w,r}
 f_{a,b,w,0}\, \bio a d^{-(a,b)}) \and \cr
&  L_{s}^0 L_{r}^0= L_{p_s p_r}^0= L_{r}^0 L_{s}^0 \for
L_r^0=L_{p_r}^0,\  r,s\in O^*.
}
\eqnu
$$
The calculation of $L_r^0$ is not complicated
(see Proposition \ref\leading) and results in

\proclaim{Proposition}% 3.3 }
The operators
$$
\eqalign{
L_r^0 \  = \
 \sum_{w\in W}& \Bigl( \prod _{\al \in \la(b_r),k\ge 0 }
{1-X_{w(\al^\vee)}q_{\al}\de^{k} \over
1-X_{w(\al^\vee)}\de^{k}} \cr
&\prod _{\al \in \la(b_r),k> 0 }
{1-X_{w(\al^\vee)}^{-1}q_{\al}^{-1}\de^{k} \over
1-X_{w(\al^\vee)}^{-1}\de^{k}} \Bigr)\, w(-\bio_r)
}
%\eqno(3.5)
\eqnu
$$
are pairwise commutative and $W$-invariant.
\endproclaim
\proofbox

The last application will be the
following parametric deformation of
the above $G$ and $\g$.
Let us introduce the second system of the (same) groups
$ W'\subset(W^b)'\subset\w'$
 and the
corresponding $\{X',\de',t'\}$, assuming  that $\w,\{X\}$
commute with  $\w',\{X'\}$. We set
$$
\eqalign{
&F_{\tal}   = { G_{\tal}+
(1-q_{\tal})(X_{\tal^\vee}' -1)^{-1}s_{\tal}
\over
1+ (1-q_{\tal})(X_{\tal^\vee}' -1)^{-1} },\
 \tal\in R^a, \cr
&\f_{\al} \equal\cdots F_{[\al,2]}F_{[\al,1]}
F_{\al} F_{[\al,-1]} F_{[\al,-2]}\cdots \ .
}
\eqnu %&(2.17)
$$
\label\F\eqnum*
Then $F$ is an
 invariant affine closed R-matrix
with  respect to the diagonal action
of $W^b$  ($\hw(\hw)'$ instead of $\hw$).
Moreover it is unitary ($F_{-\tal}=F_{\tal}^{-1}$).
Hence $\{\f_{\al}\}$  is an invariant unitary R-matrix
as well as its affine extension
$\{\f_{\tal}\}$. The latter is defined
  due to Theorem \ref\extended
relative to the diagonal action of $(W^b)^\iota$.
The
definition of the proper algebra \FF\ containing
$\f_{\tal}$ is a direct version of Definition 4.1
for two sets of variables
(see [C3], Proposition 2.5). Following
 Definition
\ref\qkz  we can use this $\f$ to
introduce the  {\it affine difference-elliptic
quantum Knizhnik-Zamolodchikov equation},
which is  connected with the eigenvalue problem
for the above operators $\{L_p\}$ (cf. [C3],
Theorem 3.5).

%
%
%
%      REFERENCES
%
%
%
%\vskip 15pt
\AuthorRefNames [BGG]
\references
%\medskip
%\ninerm
%\baselineskip=11pt %!

[B]
\name{N. Bourbaki},
{\it Groupes et alg\`ebres de Lie}, Ch. {\bf 4--6},
Hermann, Paris (1969).

[C1]
\name{I.Cherednik},
{  Double affine Hecke algebras,
Knizhnik- Za\-mo\-lod\-chi\-kov equa\-tions, and Mac\-do\-nald's
ope\-ra\-tors},
IMRN (Duke M.J.) {  9} (1992), 171--180.

[C2]
\bibline,{ Double affine Hecke algebras and  Macdonald's
conjectures},
Annals of Mathematics (1994).

[C3]
\bibline,
{ El\-lip\-tic quan\-tum many- body prob\-lem and doub\-le af\-fine
Knizh\-nik- Za\-mo\-lod\-chi\-kov equa\-tion},
Commun. Math. Phys. (1994).

[C4]
\bibline,{  Quantum Knizhnik- Za\-mo\-lod\-chi\-kov
equa\-tions and affine
root systems}, Commun. Math. Phys. {  150} (1992), 109--136.

[C5]
\bibline,
{ Integration of quantum many- body problems by affine
Knizhnik--Za\-mo\-lod\-chi\-kov equations},
Pre\-print RIMS--{  776} (1991),
(Advances in Math.(1994)).

[DJO]
\name{ C.F. Dunkl},  \name{ M.F.E. de Jeu}, and
 \name{ E.M. Opdam},
{Singular polynomials for finite reflection groups},
Preprint (1994).

[FR]
\name{ I. Frenkel}, and \name{N. Reshetikhin},
{Quantum affine algebras and holonomic difference equations},
Commun. Math. Phys. {146} (1992), 1--60.

[He]
\name{G.J. Heckman},
{  An elementary approach to the hypergeometric shift operators of
Opdam}, Invent.Math. {  103} (1991), 341--350.
omp. Math. {  64} (1987), 329--352.
v.Math.{  70} (1988), 156--236.

[KL]
\name{D. Kazhdan}, and \name{ G. Lusztig},
{  Proof of the Deligne-Langlands conjecture for Hecke algebras},
Invent.Math. {  87}(1987), 153--215.

[KK]
\name{B. Kostant}, and \name{ S. Kumar},
{  T-Equivariant K-theory of generalized flag varieties,}
J. Diff. Geometry{  32}(1990), 549--603.

[M1]
\name{I.G. Macdonald}, {  A new class of symmetric functions },
Publ.I.R.M.A., Strasbourg, Actes 20-e Seminaire Lotharingen,
(1988), 131--171 .

[M2]
\bibline, {  Orthogonal polynomials associated with root
systems},Preprint(1988).

[N]
\name{M. Noumi},
{  A realization of Macdonald's symmetric
polynomials om quantum homogeneous spaces},
in : Proceedings of the 21st international conference
on differential geometry methods in theoretical physics,
Tian, China (1992) (to appear).

[O]
\name{E.M. Opdam},
{  Some applications of hypergeometric shift
operators}, Invent.Math.{  98} (1989), 1--18.

[R]
\name{S.N.M. Ruijsenaars},
{Complete integrability of relativistic Calogero-Moser
systems and elliptic function identities}, Communs Math. Phys.
{ 110} (1987), 191--213.

[V]
\name{D-N.Verma},
{  The role of affine Weyl groups in the
representation theory of algebraic Chevalley groups
and their Lie algebras},
Lie groups and their representations (Proceedings of
the Summer School on Group Representations),
Budapest (1971),   653--705.

\endreferences

\bye